\documentclass[aip,jcp, reprint]{revtex4-1}

\usepackage{graphics}
\usepackage[dvips]{color}
\usepackage[T1]{fontenc}
\usepackage[latin1]{inputenc}
\usepackage{graphicx}
\usepackage{amsmath,amssymb}
\usepackage{rotating}
\usepackage{epsfig}
\usepackage{pstricks}

\usepackage{dcolumn}
\usepackage{bm}

\usepackage{ulem} 
\input epsf.sty

\newcommand{\mb}{\mbox}

\newcommand{\vektor}[1]{{\vec{#1}}}

\newcommand{\kb}{k_{\mbox{\tiny B}}}
\newcommand{\dx}{\mathrm{d}}

\newcommand{\indexHB}{{\mbox{\scriptsize H}}}

\newcommand{\indexDH}{{\mbox{\scriptsize DH}}}
\newcommand{\indexHA}{{\mbox{\scriptsize HA}}}
\newcommand{\indexHC}{{\mbox{\scriptsize HC}}}

\newcommand{\indexG}{{\mbox{\scriptsize G}}}
\newcommand{\indexK}{{\mbox{\scriptsize K}}}
\newcommand{\indexC}{{\mbox{\scriptsize H}}}
\newcommand{\indexAlg}{{\mbox{\scriptsize alg}}}
\newcommand{\indexRe}{{\mbox{\scriptsize re}}}
\newcommand{\indexEv}{{\mbox{\scriptsize ev}}}
\newcommand{\indexOU}{{\mbox{\scriptsize OU}}}
\newcommand{\indexSim}{{\mbox{\scriptsize sim}}} 

\DeclareMathOperator{\erfc}{erfc}
\DeclareMathOperator{\erf}{erf}

\newcommand{\power}{{\,2}}
\newcommand{\stern}{{\hspace*{0.3ex}*}}
\newcommand{\kbT}{{\kb
T_{\scriptsize\mb{room}}}}

\begin{document}

\title{Brownian Dynamics Simulations with Hard-Body Interactions: Spherical Particles}

\author{Hans Behringer}
\email{behringh@uni-mainz.de}
\affiliation{Johannes Gutenberg-Universit\"at Mainz, Institut f\"ur Physik, Staudinger Weg 7, D-55128 Mainz, Germany}
\author{Ralf Eichhorn}
\email{eichhorn@nordita.org}
\affiliation{Nordic Institute for Theoretical Physics (NORDITA), Royal Institute of Technology and Stockholm University,
Roslagstullsbacken 23, 106 91 Stockholm, Sweden}

\begin{abstract}

A novel approach to account for hard-body interactions in (overdamped)
Brownian dynamics simulations is proposed for systems with
non-vanishing force fields. The scheme exploits the analytically known
transition probability for a Brownian particle on a one-dimensional
half-line.  The motion of a Brownian particle is decomposed into a
component that is affected by hard-body interactions and into
components that are unaffected. The hard-body interactions are
incorporated by replacing the 'affected' component of motion by the
evolution on a half-line. It is discussed under which circumstances
this approach is justified. In particular, the algorithm is developed
and formulated for systems with space-fixed obstacles and for systems
comprising spherical particles.  The validity and justification of the
algorithm is investigated numerically by looking at exemplary model
systems of soft matter, namely at colloids in flow fields and at
protein interactions. Furthermore, a thorough discussion of properties
of other heuristic algorithms is carried out.
\end{abstract}

\maketitle

\section{Introduction}

Many physical processes in soft matter systems, like colloids
\cite{Jones2002,Dhont2003,Oettinger1996}, polymers \cite{Doi1986} or
biomolecular systems \cite{Jackson2006}, can be modeled in terms of
Langevin equations where the influence of the solvent is captured by
friction and random forces.
\cite{Gardiner1983,Risken1984,vanKampen1987,Honerkamp1990} The
separation of the respective length and time scales of the solvent
particles and the constituent particles of soft matter typically
allows one to use the high friction limit where the velocities of the
latter particles are strongly damped and inertia effects are thus
negligible (overdamped Langevin equations).  Another common
idealization in the modeling of mesoscopic physical systems within the
framework of Langevin equations consists in representing the extremely
short-ranged and strong repulsive contact forces between particles or
between particles and space-fixed obstacles or walls by hard-body
interactions.  Important examples for physical processes which are
investigated using these limits include the dynamical behavior of
polymers and colloids in dilute as well as concentrated solutions, the
self-assembly of biomolecules into biological structures like
membranes or protein aggregates, the motion of molecules in
biomolecular systems and the migration of macromolecules in
microfluidic channels \cite{Nguyen2006}.

To unravel the properties of such physical processes, a thorough
investigation of the particles' trajectories is often
indispensable. To name just two examples, the knowledge of the typical
trajectories is crucial to analyze association pathways of
complex-forming proteins or the migration mechanisms
of biomolecules in microfluidic channels.  As in most
cases an analytical treatment of the Langevin equations of motion is
not possible on such detailed, trajectory-wise level, efficient
numerical integration schemes for solving these stochastic
differential equations are essential. \cite{Kloeden1999,Snook2007}
These schemes require a certain degree of regularity of the involved
force fields, \cite{Kloeden1999} so that hard-body interactions
(associated with reflective boundaries\cite{noteAbsorbing}) can not be
integrated directly due to their singular nature.  Therefore, it is
necessary to represent the effect of singular hard-body interactions
in a tractable form.  In order to avoid the numerically costly
introduction of steep (and regular) auxiliary potentials, alternative
potential-free methods seem particularly promising with respect to
computational efficiency.  Such methods are based on detecting
'unphysical configurations', i\,e. 'collisions' or 'overlaps' between
hard-body entities that arise from a regular numerical integration
step, and on some rule that specifies how this unphysical
configuration is to be corrected in order to create a physically valid
configuration.  Several heuristic methods along these lines have been
discussed in the literature, ranging from Monte Carlo type schemes,
where unphysical configurations are just rejected, to methods where
overlapping particles with hard cores are displaced along their
connection line. This is done by applying specific rules that are
based on geometric considerations or on the physics of elastic
collisions.  \cite{Cichocki1990,Schaertl1994, Strating1999} In recent
years the emergence of event-driven molecular dynamics has inspired
the development of trajectory-based heuristic 'event-driven' Brownian
dynamics schemes. \cite{Tao2006,Scala2007} These methods often lack a
thorough justification. \cite{fussnoteScala} Apart from the rejection
method \cite{Cichocki1990} and the event-driven scheme of Tao et
al. \cite{Tao2006} it is not obvious how the different methods can be
applied to driven diffusive systems with additional non-vanishing
force fields.

In this work we propose a novel method to account for hard-body
interactions in numerical Brownian dynamics simulations. The new
approach reverts to the transition probability for the time evolution
of a particle whenever a hard-body interaction has to be accounted
for. This transition probability is determined by the
Fokker-Planck equation which provides an alternative way to describe
the physics of stochastic
processes. \cite{Gardiner1983,Risken1984,vanKampen1987} Once this
transition probability is known, a physical trajectory of a particle
can be directly generated. However, for many systems this transition
probability is not known analytically. In this paper we will argue and
discuss how the analytical solution of the Fokker Planck equation for
the one-dimensional half-line can be used to account for hard-body
interactions in general, more-dimensional systems. The new algorithm
has the advantage that it can be applied to systems with spatially
varying force-fields and can be used independently of the applied
numerical integration scheme. In addition, the effects of the
introduced approximations due to applying the transition probability
for a system on a half-line can be critically assessed so that these
approximations can be justified and numerically controlled.

The article is organized as follows. In Sec. \ref{sec:Brownian} the
basics of Brownian dynamics simulations are recapitulated. Then the
problem of accounting for hard-body interactions in Brownian dynamics
is discussed and the basic idea of our novel approach to incorporate them is outlined.
In this context other proposed heuristic methods are
discussed in Sec. \ref{sec:hardWallHeuristics}. Section
\ref{sec:hardWall} is then devoted to the development of the new
algorithm for space-fixed hard walls. The algorithm is next
generalized to systems with hard spheres in Sec.
\ref{sec:hardSpheres}. The validity, quality and efficiency of the methods are
scrutinized in Sec. \ref{sec:exemplaryModels} by numerically
investigating different exemplary physical situations with
non-vanishing force fields. A brief account of some of the results of
this article was already published elsewhere.\cite{Behringer2011}

\section{Hard-body interactions in Brownian Dynamics}
\label{sec:Brownian}

\subsection{Basics of Brownian Dynamics simulations}

Consider a Brownian particle in a solution and in a force field
$\vektor{F}=\vektor{F}(\vektor{r},t)$ at temperature $T$. Apart from
the deterministic force $\vektor{F}$ due to external potentials,
non-conservative forces or inter-particle interactions, the Brownian
particle experiences the influence of the medium.  One influence is
related to the friction which is determined by Stokes' friction force
$-\eta \dot{\vektor{r}}$ with friction coefficient $\eta$ and particle
velocity $\dot{\vektor{r}}$. In addition, the medium influences the
particle by random collisions through its constituents.  This thermal
heat bath is represented by a random force $\vektor {\xi}$ with the
statistical properties of Gaussian random variables:
$\langle\xi^\alpha(t) \xi^\beta(s)\rangle = \delta^{\alpha \beta}
\delta(t-s)$ with $\alpha$ and $\beta$ denoting Cartesian
components. In the strong friction limit, for which inertia effects
are negligible, the evolution of the particle is then governed by the
Langevin equation
\begin{equation}
\label{eq:basicLangevin}
\dot{\vektor{r}}(t) = \frac{\vektor{F}(\vektor{r}(t),t)}{\eta} + \sqrt{2D} \vektor{\xi}(t),
\end{equation}
where the strength of the fluctuating force is related to the
diffusion constant $D = \kb T/\eta$ (overdamped Langevin dynamics,
Brownian dynamics). \cite{Risken1984,Oettinger1996,Snook2007} Note
that the assumption of the strong friction limit is well satisfied for
biological, colloidal or polymeric systems.
\cite{Doi1986,Jackson2006}

The Langevin evolution equation can be used to obtain the trajectory
of the particle whose physical contents can then be investigated.  In
this context the Langevin equation is best formulated in terms of
differentials,
\begin{equation}
\label{eq:langevinWithWiener}
\dx \vektor{r} = \vektor{v} \dx t + \sqrt{2D} \dx \vektor{W}.
\end{equation}
Here the drift term is represented in terms of a velocity $\vektor{v}
= \vektor{F}/\eta$ and the stochastic term is related to a Wiener
process $\vektor{W}$. The Wiener process $\vektor{W}$ describes free
Brownian diffusion and is mathematically defined to be a Gaussian
stochastic process with mean $\langle {W}^\alpha(t)\rangle = 0 $ and
covariance $\langle W^\alpha(t)W^\beta(s)\rangle = \delta^{\alpha
  \beta}\min (t,s)$.  \cite{Gardiner1983,Roepstorff1991,Paul1999} In
addition, the It\^{o} Lemma $\langle \dx W^\alpha(t)\dx
W^\beta(t)\rangle = \dx t \delta^{\alpha \beta}$ holds true for its
differential. Note that the incremental Wiener process $\dx
\vektor{W}(t)$ cannot be written in the form $\vektor{w}(t)\dx t$ due
to the non-differentiability of the trajectory of a freely diffusing
particle. \cite{Gardiner1983,Roepstorff1991,Paul1999}

The numerical integration of (\ref{eq:langevinWithWiener}) will
generate a discrete approximation of the trajectory. To this end a
discrete 'incremental' displacement corresponding to
(\ref{eq:langevinWithWiener}) is generated for a finite time step from
the current physical position of the particle.  The most prominent
integration scheme for such Brownian dynamics simulations is the
Euler algorithm (see Sec. \ref{sec:hardWallHeuristics}).  Apart from
single step algorithms like the Euler method, multistep integration
schemes such as the Heun algorithm have been developed that
additionally incorporate the information from configurations for
intermediate steps (more details about the numerical integration of
stochastic differential equations, in particular on necessary
conditions on the regularity of the involved forces, can be found in
the literature
\cite{Greiner1988,Oettinger1996,Kloeden1999,Snook2007}).

One crucial point in carrying out a numerical integration of
(\ref{eq:langevinWithWiener}) is the choice of an appropriate size of
the time step between two successive times of the discrete
trajectory. For a non-zero and spatially varying force field
$\vektor{F}$, the time step has to be small enough so that the force is
virtually constant on the scale that is associated with the typically
proposed spatial displacement.  For systems whose time evolution is
usually modelled within a Brownian dynamics approach, however, the
corresponding force fields are in general relatively smooth. This
allows the use of fairly large integration time steps.

\subsection{Algorithm for accounting hard-body interactions}
\label{sec:algo}

For systems with hard-body interactions the direct numerical
integration of the unmodified Langevin equations over a given time
step can result in unphysical configurations where the particles are
displaced into the wall, for example, or where hard-core particles
penetrate each other. Such configurations,
which are obtained from the direct integration
of the Langevin equation but are not consistent with hard-body
interactions will be called 'unphysical' in the following.
As the hard-body interactions are singular in
nature they cannot directly be incorporated into the integration
scheme. One possibility is given by a representation in terms of
regular potentials. This approach, however, has the drawback, that the
auxiliary potentials have to be rather steep and therefore the
acceptable time step for the numerical integration has to be very
small rendering the simulation extremely time-consuming.  Thus other more
effective potential-free ways to account for hard-body interactions
are desirable. \cite{noteSectionHeuristic}

The method we propose here is based on the approach to consider the physics of a
diffusing particle in a force field $\vektor{F}$ by directly looking at
the transition probability $p(\vektor{r},t;\vektor{r}_0,t_0)$ to find
the particle at position $\vektor{r}$ at time $t$ provided it was at
position $\vektor{r}_0$ at time $t_0 < t$. This (normalized)
probability is governed by the Fokker-Planck equation\cite{Gardiner1983,Risken1984,vanKampen1987}
\begin{equation}
\label{eq:basicFokkerPlanck}
\frac{\partial}{\partial t} p(\vektor{r},t;\vektor{r}_0,t_0) = \left(D \Delta_{\vektor{r}}  - \vektor{\nabla}_{\vektor{r}} \cdot \frac{\vektor{F}(\vektor{r},t)}{\eta}\right)p(\vektor{r},t;\vektor{r}_0,t_0). 
\end{equation}
To solve this partial differential equation boundary conditions for
the physical domain have to be provided. The presence of hard walls at
a boundary with normal vector $\vektor{n}$ corresponds to reflective
boundary conditions for which the normal projection
$\vektor{n}\cdot\vektor{j}$ of the probability flux
\begin{equation}
  \vektor{j}(\vektor{r},t;\vektor{r}_0,t_0) = -\left(D\vektor{\nabla}_{\vektor{r}}  - \frac{\vektor{F}(\vektor{r},t)}{\eta}\right) p(\vektor{r},t;\vektor{r}_0,t_0)
\end{equation}
vanishes at the boundary.

In general, a realization for the displacement $\vektor{r} -
\vektor{r}_0$ of a Brownian particle after elapsed time $t-t_0$ can be
directly generated from the transition probability
$p(\vektor{r},t;\vektor{r}_0,t_0)$.  Suppose now that this probability
is known in the vicinity of the hard surface and that the numerical
integration using the unmodified Langevin equation (without hard
walls) has led to an unphysical position $\vektor{r}$ for the particle
that was initially at $\vektor{r}_0$. Instead of applying a modified
integration step to the Langevin equation, a new physical position
$\vektor{r}^\stern$ for the elapsed time $t$ is directly generated
from that part of the transition probability
$p(\vektor{r}^\stern,t;\vektor{r}_0,0)$ that represents the particle's
encounter with the wall.  This recourse to the transition probability
whenever a hard-body interaction has to be accounted for forms the
heart of our proposed algorithm.  In the following we will demonstrate
that the transition probability $p(\vektor{r},t;\vektor{r}_0,t_0)$ for
spherical particles can indeed be decomposed in a mathematically
rigorous way into a 'free' part (without hard-body interactions) and a
part that stems from the reflective hard boundary. This decomposition
is based on the solution of the Fokker-Planck equation on a
one-dimensional half-line with a reflecting boundary at the origin.

\subsection{Exact one-dimensional algorithm} 

For the particular example of a one-dimensional system, where the
position $q$ is restricted to the half-line,
$[0,\infty [ $ and is subject to the constant force $f$ the
    Fokker-Planck equation reads
\begin{equation}
\label{eq:onedimFokkerPlanck}
  \frac{\partial}{\partial t}p =  D_q\frac{\partial^2}{\partial q^2} p -\frac{f}{\eta}\frac{\partial}{\partial q} p. 
\end{equation}
The transition probability can be calculated analytically for a
reflective boundary at $q=0$.
\cite{Smoluchowski1916,Chandrasekhar1943} Using the
notation $v := f/\eta$ and setting $t_0=0$ without loss of generality the solution for the transition
probability $p(q, t;q_0) \equiv p(q,t;q_0,t_0 = 0)$ is given by
\begin{equation}
\label{eq:solutionOnedimFokkerPlanck}
  p(q, t; q_0) = p_1(q, t;q_0) + p_2(q, t;q_0) +  p_3(q, t;q_0)
\end{equation}  
with the contributions
\begin{subequations}
\label{eq:p1dimForm}
\begin{eqnarray}
\label{eq:p1Form}
p_1(q, t; q_0) &=& \frac{1}{\sqrt{4\pi D_qt}} \exp\left(- \frac{(q-q_0-vt)^2}{4D_qt}\right),\\
p_2(q, t; q_0) &=& \frac{\exp\left(-\frac{vq_0}{D_q}\right)}{\sqrt{4\pi D_qt}} \exp\left(- \frac{(q+q_0-vt)^2}{4D_qt}\right)
\end{eqnarray}
and
\begin{eqnarray}
\label{eq:p3Form}
p_3(q, t; q_0) &=& -\frac{v}{2D_q} \exp\left(\frac{vq}{D_q}\right) \erfc\left( \frac{q+q_0+vt}{\sqrt{4D_qt}}\right).
\end{eqnarray}
\end{subequations}
Each $p_i$ satisfies the Fokker-Planck equation
(\ref{eq:onedimFokkerPlanck}) individually, their sum then satisfies
the initial condition $p(q, t=0; q_0) = \delta(q-q_0)$, the
reflective boundary condition
$j(q,t;q_0)|_{q=0} = -(D_q\frac{\partial}{\partial x} p - v p)|_{q=0} = 0$,
and is normalized on the half-line $[0,\infty [ $. The different parts
have the following intuitive interpretation. The first term $p_1$ is
the probability distribution that corresponds to unrestricted
diffusion in a constant force field and represents the contributions
from particles that do not encounter the wall at $q=0$ during the
propagation over the time $t$. The propagation of such particles is
already correctly captured by the 'free' Langevin equation $\dx q = v \dx t +
\sqrt{2D_q} \dx W$ corresponding to (\ref{eq:onedimFokkerPlanck})
without boundary at $q=0$. The
second and third terms contain the contributions from particles that
encounter the wall during the propagation time $t$. These
contributions $p_2$ and $p_3$ are obviously not correctly contained in
the associated Langevin equation without further consideration of the
reflective boundary condition at $q=0$.

\subsubsection{The algorithm}
It is now obvious how physical positions that are consistent with a
reflective boundary at $q = 0$ can be generated:
\begin{enumerate}
\item[1.]
Starting at $q_0>0$, suggest a new
position $q$ of the Brownian particle (with diffusion constant $D_q$)
after time step $\dx t$ by some
given numerical integration scheme for the Langevin equation
(without further consideration of the hard wall).
\item[2\,a.]
If the proposed $q$ is in the physical domain, i.e. $q>0$,
it is accepted as the new particle position.
\item[2\,b.]
If the proposed position is unphysical, i.e. $q<0$, a new
position is generated directly from the transition probability
(see Appendix \ref{sec:generateRandomQ})
\begin{eqnarray}
\label{eq:distributionOutwardWall}
  p_\indexHB(q,q_0,\dx t,v, D_q) = \frac{p_2(q) + p_3 (q)}{ \int_0^\infty\dx q \left(p_2(q) + p_3 (q) \right)}.
\end{eqnarray}
\end{enumerate}
Obviously, the end positions which are directly accepted from
the integration step of the Langevin equation make up the
contribution $p_1$ in the exact solution
(\ref{eq:solutionOnedimFokkerPlanck}) of the transition probability.
The distribution $p_\indexHB$ represents the (normalized) part
$p_2+p_3$ in the analytical solution
(\ref{eq:solutionOnedimFokkerPlanck}) capturing the physics of
particles that encounter the wall at least once. See Appendix
\ref{sec:generateRandomQ} on how to generate $q$ from
(\ref{eq:distributionOutwardWall}) in practice.

\subsubsection{Discussion}
The algorithm is constructed such that a stochastic particle trajectory
is generated from two contributions: Parts where the
particle does not encounter the wall at all are captured by
a standard integration step,
whereas the parts of the trajectory where wall-collisions occur are
constructed from $p_\indexHB$.
Let us look at this point from a different angle. Mathematically
speaking the algorithm uses two random variables $X_\indexG$ and
$X_\indexC$ to construct a valid end position $X$. The Gaussian variable $X_\indexG$ describes unrestricted free diffusion, can thus take on
all real numbers $\mathbb{R}$ and is distributed according to the Gaussian
distribution $p_1$ (compare (\ref{eq:p1Form})), which is normalized on $\mathbb{R}$.
$X_\indexC$ can take on only positive real numbers $\mathbb{R}_{> 0}$ and is distributed
according to (the normalized) $p_\indexC$ of relation
(\ref{eq:distributionOutwardWall}). The final random variable generated from the algorithm is of the form
\begin{equation}
\label{eq:constructPalg}
  X = X_\indexG \theta (X_\indexG) + X_\indexC \theta(-X_\indexG),
\end{equation} 
where $\theta(X)$ is the Heaviside step function.
Defining the collision probability
\begin{eqnarray}
w
& = & \int\limits_0^\infty \dx q (p_2(q)+p_3(q))
\label{eq:w} \\
& = & \int\limits_0^\infty \dx q (1-p_1(q)) = \int\limits_{-\infty}^0 \dx q p_1(q)
\nonumber \\
& = & \frac{1}{2} \erfc\left(\frac{q_0+v\dx t}{\sqrt{4D_q \dx t}}\right)
\end{eqnarray}
one can write $p_\indexC = (p_2+p_3)/w$. The
random variable $X$ associated with the algorithm has the distribution function
\begin{eqnarray}
  p_\indexAlg(q) &=& \int\limits_{-\infty}^{+\infty} \dx x_1
  \int\limits_0^\infty \dx x_2 p_\indexG(x_1)p_\indexC(x_2) \\ && \quad \times \delta(q -
  [x_1\theta(x_1) + x_2\theta(-x_1)]) \\
 &=&
  \int\limits_{-\infty}^{+\infty} \dx x_1
  \int\limits_0^\infty \dx x_2 p_1(x_1)\frac{p_2(x_2)+p_3(x_2)}{w} \\ && \quad \times \delta(q -
  [x_1\theta(x_1) + x_2\theta(-x_1)]).
\end{eqnarray}
Due to the Heaviside functions in the argument of the delta function the latter can be rewritten as 
$\delta(q -
  [x_1\theta(x_1) + x_2\theta(-x_1)]) = \theta(x_1)\delta(q -x_1) + \theta(-x_1)\delta(q-x_2)$.
The resulting two integrals finally add up to
$p_\indexAlg(q) = p_1(q)+p_2(q)+p_3(q)$ and hence the algorithm indeed
generates random variables which are distributed according to the
Smolouchowski solution (\ref{eq:solutionOnedimFokkerPlanck}).

By construction our proposed procedure for a simulation in a system
with constant force will generate a trajectory with the correct
statistics independent of the size of the used time step $\dx t$.  For
a slowly varying force the time step can be chosen small enough so
that the force stays virtually constant for the typical displacement
of the particle. For each time step the simulation is then carried out
as above with the presently acting force (see Appendix \ref{sec:OU}
for an illustration). For such an appropriately adjusted time step the
Brownian dynamics simulation will generate a trajectory that exhibits
the correct physical properties in a controllable accuracy.

For the outlined approach to work for a general physical system one
has to solve the Fokker-Planck equation for the transition probability
in the vicinity of a hard surface analytically. This is only possible
in closed form for a very limited number of problems. In the following
sections we will show in details that the solution for the
one-dimensional half-line can be used to construct an approximate
transition probability with which hard-body interactions can be
accounted for in many situations. The approximation is thereby
controlled by the size of the time step. For small time steps only
small displacements show up so that a generally shaped hard surface appears to
be virtually flat. For the normal direction the systems is then
basically restricted to a half-line whereas the motion in the
components parallel to the surface is unrestricted. This parallel
motion is decoupled from the normal motion and can thus be treated
within the usual approach whereas the one-dimensional normal motion is
dealt with by the one-dimensional approach from above.
Details will be worked out below. We note already
at this point, however, that the involved adaption of the time step
for spatially structured hard boundaries works
along the same lines as the corresponding choice to capture the
effect of spatially varying force fields.

\section{Heuristic representations of hard walls}
\label{sec:hardWallHeuristics}

Before extending the outlined approach to more complex physical systems
we discuss the quality and validity of other proposed heuristic
methods to account for hard-body interactions in a potential-free
manner.  To this end we consider a Brownian particle on the
one-dimensional half-line in a constant force field for which the
analytical solution (\ref{eq:solutionOnedimFokkerPlanck}) is at
hand. This special system will enable us to precisely identify the
shortcomings of those schemes.  In the following we will only compare
methods that can be directly extended to systems with non-vanishing
forces.

\subsection{Rejection scheme}

\begin{figure}[h!]
\begin{center}
\includegraphics[scale=0.325,angle=0]{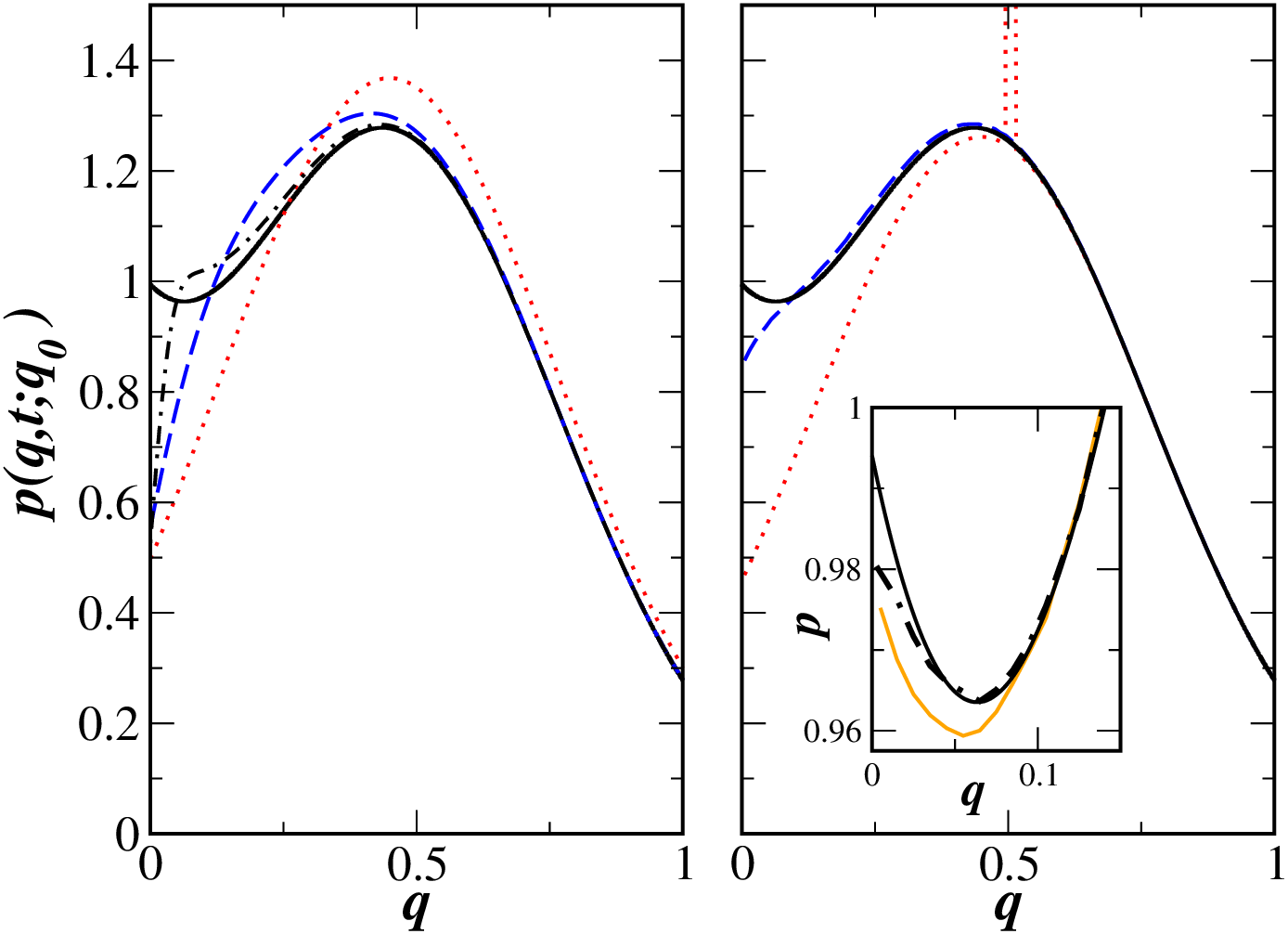}
\caption{\label{bild:onedimensionalRejection} Transition probability
  of a particle on a half-line for the parameters $v=-1.0$, $D_q=1.0$,
  initial position $q_0=0.5$ (corresponding to an encounter
  probability $w$, equation (\ref{eq:w}), of approximately 8\%) and
  elapsed time $t=0.05$. \cite{noteDimensionless} The analytical curve
  (\ref{eq:solutionOnedimFokkerPlanck}) is displayed by the solid line
  and is compared to results from numerical simulations based on the
  Euler algorithm using the rejection method without advancing time
  (left) and updating time (right). The (red) dotted curve is obtained
  with $\dx t = 0.05$.  For the rejection scheme with updating time a
  delta-like peak at the initial position emerges representing the
  probability to encounter the wall.\cite{SchaertelNote} The (blue)
  dashed and the dotted-dashed curves were obtained using $\dx t =
  0.005 = t/10$ and $\dx t = 0.0005 = t/100$, respectively. (For
  comparison the light (orange) line in the inset shows results from
  the event-driven method discussed below with $\dx t = t = 0.05$.)}
\end{center}
\end{figure}

A first possible scheme could consist in
discarding all displacements into hard walls as unphysical and
repeating the integration step until a valid displacement is
obtained. However, in doing so the influence of the contributions
$p_2$ and $p_3$ is not captured even approximately, as this scheme tries
to reconstructed the whole transition probability only from (the
renormalized) $p_1$. This results in a very poor convergence in the
limit of decreasing integration time step $\dx t$ as can be seen in
Fig. \ref{bild:onedimensionalRejection} (left panel). In particular, one would
expect that systematic deviations show up only in close proximity of
the wall. The mentioned renormalizing of $p_1$, however, leads to
deviations even far away from the wall (compare Fig.
\ref{bild:onedimensionalRejection}, left panel).

The Monte Carlo inspired rejection scheme models the effect of hard
body interactions by dismissing a proposed propagation that leads to
an unphysical configuration, but still advancing time. \cite{Cichocki1990}
The part $p_1$ in the probability distribution
(\ref{eq:solutionOnedimFokkerPlanck}) is correctly reproduced for a
time step $\dx t$, but the contributions from $p_2$ and $p_3$ are
replaced by a delta function situated at the initial position $q_0$
with a weight given by the collision probability $w$ (compare(\ref{eq:w}),
see also Fig.~\ref{bild:onedimensionalRejection} (right panel) and
Fig.~\ref{bild:onedimensionalHeuristic}).
The one-step distribution function of the rejection scheme thus reads
\begin{equation}
\label{eq:pReject}
p_{\indexRe}(q) =
p_1(q) 
+ \frac{1}{2} \erfc \left(\frac{q_0+v\dx t}{\sqrt{4D_q \dx t}}\right) \delta(q-q_0).
\end{equation}
In this way the
relative weight of the contributions $p_1$ and $p_2+p_3$ is correctly
captured, although the influence of encounter events is not
'explicitly' worked in. In addition, the reflective
boundary condition at $q=0$ is violated. Instead of a vanishing probability flux one has 
\begin{equation}
j_\indexRe|_{q=0} =  -\frac{ \exp\left(- \frac{(q_0+v\dx t)^2}{4D_q\dx t}\right)}{\sqrt{4\pi D_q\dx t}} \frac{q_0-v \dx t}{2\dx t}.
\end{equation}

In practise a finite time step will give acceptable results for 
large initial separations from the wall where the
typical displacement will not lead to an encounter. Close to the wall
where $\sqrt{2D_q\dx t}/q_0$ is considerably larger than zero, however,
systematic deviations show up.  For illustration we give (the
asymptotic series of) the mean value $\langle q \rangle_{\indexRe}$ of
(\ref{eq:pReject})
\begin{equation}
\langle q \rangle_{\indexRe} = \langle q \rangle - \frac{ \mbox{e}^{- \frac{q_0^2}{4D_q \dx t}}}{\sqrt{4\pi D_q \dx t}}\left(\mbox{e}^{-\frac{vq_0}{2D_q}} 2D_q\dx t + \mathcal{O}(\dx t^2)\right)
\end{equation}
where $\langle q \rangle $ is the mean of the exact solution
(\ref{eq:solutionOnedimFokkerPlanck}).  (In
Refs. \onlinecite{Cichocki1990} and \onlinecite{Scala2007} the
convergence of the rejection method in the limit of vanishing $\dx t$
was discussed for systems without deterministic forces). Figure
\ref{bild:onedimensionalRejection} (right panel) illustrates the influence of the
size of the time step on the convergence behavior. The chosen
parameters correspond to a situation where the probability to
encounter the wall is already moderately small. Even then rather small
time steps have to be used to reduce the systematic deviations for end
positions close to the wall. 

We note that a modification of the rejection scheme was proposed in
Ref. \onlinecite{Schmidt2008} for many-particle systems. There the
amount for which time is advanced during one step is related to the
fraction of proposed displacements without unphysical hard-core
overlaps compared to displacements with overlaps (rejection with
'intrinsic' clock).  For just one particle, this scheme corresponds to
the rejection method without advancing time, which has been
demonstrated above (see also Fig.
\ref{bild:onedimensionalRejection}, left panel) to exhibit poor
convergence.

\subsection{Event-driven methods}

The event-driven scheme of Tao et
al. \cite{Tao2006} attempts to capture the influence of encounter
events by using an auxiliary path for the propagating particle.  For
an integration step where an overlap is detected the singular
trajectory of a Brownian particle is replaced by a regular path whose
parameterization is obtained from the used integration scheme.  To
keep the discussion clear we consider the Euler algorithm where the
new particle position $q(t+\dx t)$ is given
by
\begin{equation}
q(t+\dx t) = q(t) + v(t)\dx t +
\sqrt{2D_q \dx t} \, G
\end{equation}
with $G$ being a random Gaussian variable of zero mean and unit
variance. Suppose the particle starts from a position $q(t) = q_0$
close to the wall and its end position is in the wall. The regular
auxiliary path is then parameterized by $q(t + \lambda\dx t) = q_0 +
v\lambda\dx t + \sqrt{2D_q\dx t} \sqrt{\lambda} G$ with $\lambda \in
[0,1]$ and fixed realization $G$. From this path a 'collision
time' $\lambda_0$ with the wall can be calculated (note that
$\lambda_0$ itself is a random variable). Then the particle is
propagated along the path to the collision point corresponding to an
elapsed time $\lambda_0 \dx t$. Starting at the wall the particle is
then propagated the remaining time $\dx \tau = \dx t-\lambda_0\dx t$
with a new realization of the Gaussian random variable such that the
initial part of the path leads away from the surface and a new
physical end position is obtained.  This approach obviously captures
the influence of encounter events, however, the contributions from
$p_2$ and $p_3$ are not completely collected.  Suppose, for instance,
that the deterministic force is pointing away from the
wall, i.e. $v>0$. For a negative Gaussian variable $G$ the path for
the propagation for the remaining time $\dx \tau$ starting directly at
the wall has an initial part which will always lead into the wall
although it might produce a valid end-position. Such a path has to be
rejected. A physical end position $q\in [0,\infty[$ is therefore only
generated if the sign of the Gaussian random variable $G$ is
positive. The physical end-position is therefore systematically
shifted away from the wall by an amount of $\mathcal{O}(v\dx\tau)$. 
To avoid this systematic shift one can relax the
conditions on the random Gaussian variable and allow all paths
that lead to valid end-positions. We use this latter variant for
the discussions throughout the present paper. In addition, the event-driven approach
suffers from the serious drawback that the sampling of a
probability distribution directly from trajectories never can
generate the negative contributions of $p_3$ for forces
$v>0$. Hence, a procedure that tries to incorporate the presence
of a hard wall into a scheme that generates (discretized and
regular approximations of the) trajectories will not correctly
capture the contributions of $p_3$.

\begin{figure}[h!]
\begin{center}
\includegraphics[scale=0.325,angle=0]{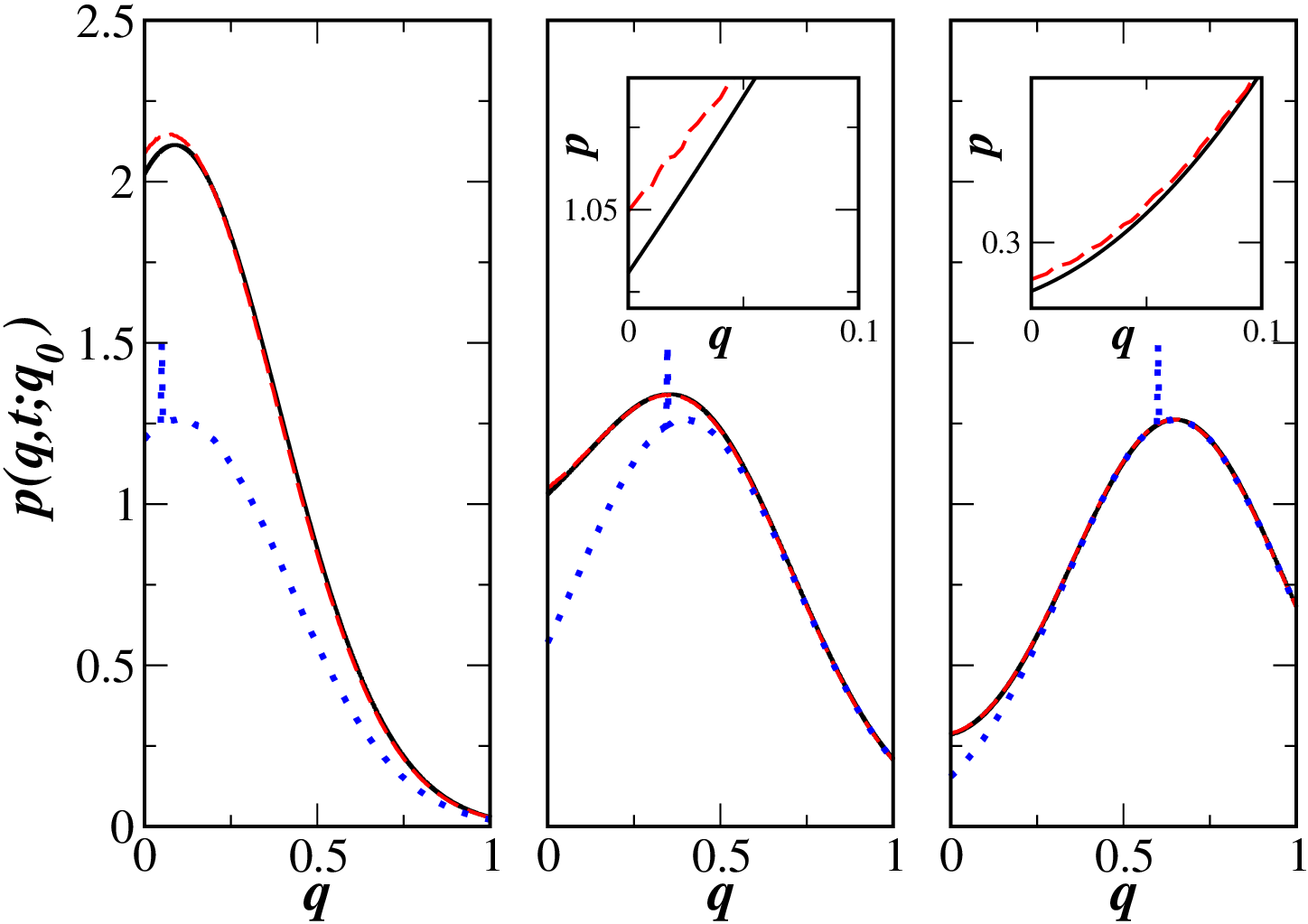}
\caption{\label{bild:onedimensionalHeuristic} Transition probability
  of a particle on the half-line for $v=1.0$, $D_q=1.0$ and elapsed time
  $t=0.05$.\cite{noteDimensionless} Three different initial positions
  $q_0$ are shown, namely, from left to right, $q_0 = 0.05$
  (corresponding to a probability $w$ of approximately 38\% to encounter
  the wall), $q_0 = 0.35$ (10\%) and $q_0 = 0.6$ (2\%). The solid
  curves depict the analytical solutions
  (\ref{eq:solutionOnedimFokkerPlanck}). The numerical results are
  obtained by one integration step of the Euler scheme. The
  (red) dashed curves correspond to the event-driven method. For comparison
  the (blue) dotted line represents the results from the rejection method. In
  these curves a delta-like peak shows up at the initial position
  $q_0$ representing the particles that encounter the wall, and are
  thus not displaced in the rejection method (indicated by a
  spike). The insets show the deviation of the results from the
  event-driven scheme close to the wall, each with the same
  magnification.}
\end{center}
\end{figure}

Figure \ref{bild:onedimensionalHeuristic} displays the analytical
solution (\ref{eq:solutionOnedimFokkerPlanck}) of the Fokker-Plank
equation (\ref{eq:onedimFokkerPlanck}) together with transition
probabilities obtained from the event-driven scheme for
different initial separations $q_0$ from the wall. As expected, the
systematic deviations are reduced for increasing $q_0$ for which the
encounter probability is diminishing as well.

In principle the one-step distribution function for the event-driven
scheme can be constructed analogously to (\ref{eq:constructPalg}) from
$ X = X_\indexG \theta (X_\indexG) + X_\indexK \theta(-X_\indexG)$
where $X_\indexK$ represents the random variable generated in case of
an encounter. This variable is distributed according to
\begin{equation}
  p_\indexK(x;\dx \tau) = \frac{\exp(- \frac{(x-v\dx \tau)^2}{4D_q\dx \tau})}{\sqrt{\pi D_q\dx \tau}\left( 1 + \erf\left(\sqrt{\frac{\dx \tau}{4D_q}v}\right)\right)}, \quad x>0.
\end{equation}  
However, the time $\dx \tau$ depends on the value of $X_\indexG<0$ of the first integration attempt and is determined by
\begin{eqnarray}
\label{eq:tauGl1}
X_\indexG &=& q_0 + v \dx t + \sqrt{2D_q\dx t} \, G,\\
\label{eq:tauGl2}
0 &=& q_0 + v (\dx t - \dx \tau) + \sqrt{2D_q (\dx t - \dx \tau)} \, G,
\end{eqnarray}
from which $G$ can be eliminated and the physical $\dx \tau \in [0,\dx t]$ be computed. The resulting distribution is thus 
\begin{equation}
\label{eq:constructProbEvent}
p_\indexEv(q) = p_1(q) + \int\limits_{-\infty}^0\dx x \,p_\indexK(q;\dx \tau(x))p_\indexG(x) \, .
\end{equation}
This expression, however, cannot be evaluated in
closed form. Recall that the event-driven scheme produces a
distribution of remaining propagation times $\dx \tau \in [0,\dx
  t]$. Due to relations (\ref{eq:tauGl1}) and (\ref{eq:tauGl2}) this
can be expressed in terms of unphysical $x \in ]-\infty, 0]$ as is
done in the integral in (\ref{eq:constructProbEvent}).

We already noted that the event-driven scheme includes effects due to
encounter events with the wall. In particular, all events that
encounter the wall precisely once are correctly captured whereas those
with several encounters are approximated by trajectories with one
encounter only. This observation readily suggests a refinement of the
event-driven scheme. Once the particle is propagated to the first
collision point the event-driven scheme is applied to the remaining
time interval so that it is either propagated directly into the valid
region or again a further collision point and collision time have to
be determined. This approach is then iteratively applied till the
particle is propagated for the total time step $\dx t$. From numerical
investigations (not shown in this article) we conclude that this
iterative event-driven scheme indeed leads to correct transition
probabilities for forces into the wall. For forces away from
the wall, however, the procedure is not correct as the negative
contribution from $p_3$ cannot be captured in this trajectory-based
method.

We note at this point that Lamm and Schulten \cite{Lamm1981} developed
also a scheme which successively takes overlap events into account. If
an overlap is detected a further displacement away from the wall is
determined with a remaining propagation time that is sampled with the help of
an approximation of the collision time distribution. If this still
leads to an overlap a third step is generated according to
(\ref{eq:p3Form}) approximated by a Boltzmann distribution. For
details see Ref. \onlinecite{Lamm1981}.

\subsection{Supplementary remarks}

\begin{figure}[h!]
\begin{center}
\includegraphics[scale=0.325,angle=0]{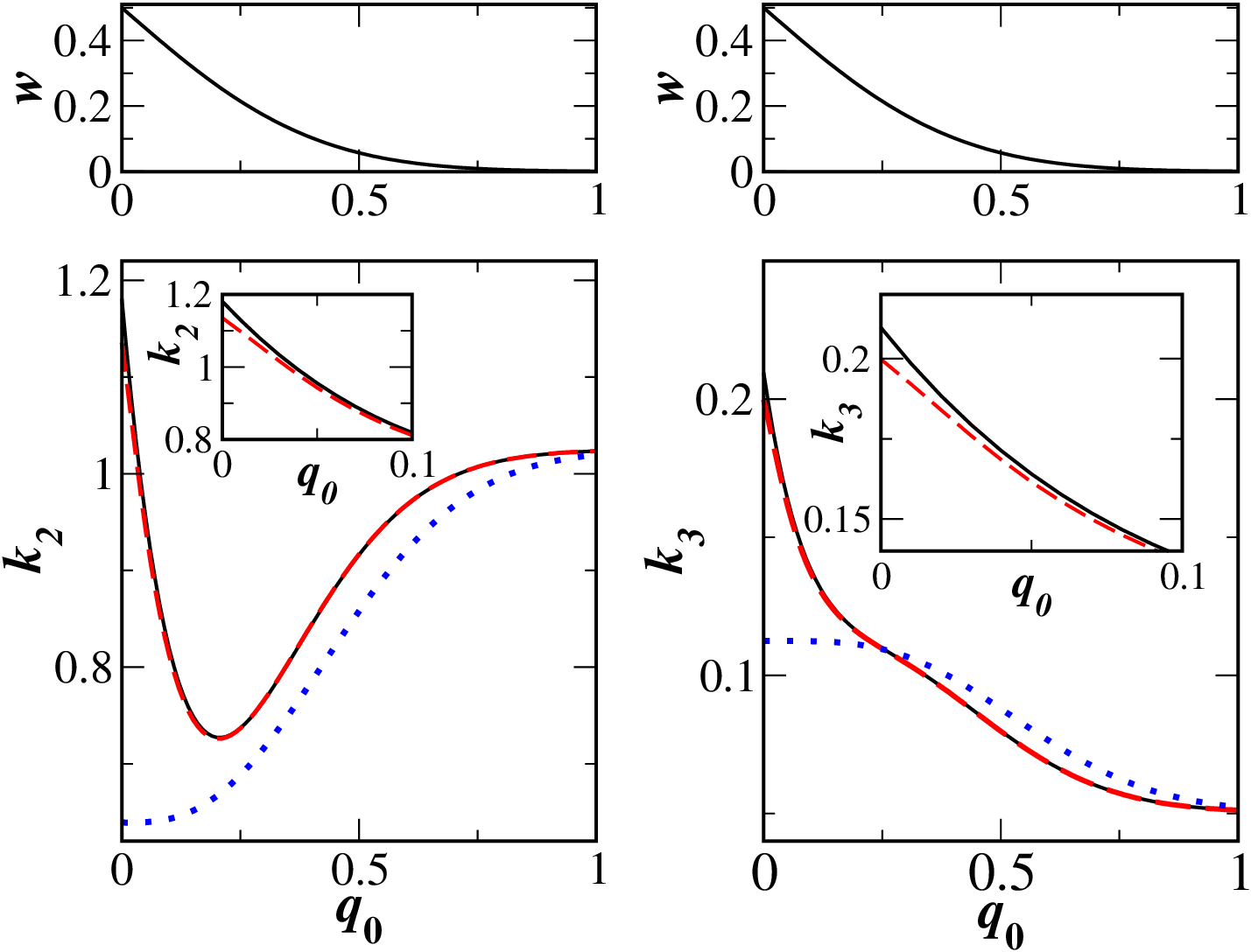}
\caption{\label{bild:dispersionAtWall} Second and third
  ('finite-time') coefficient $k_2(\dx t; q_0)$ (lower left panel) and
  $k_3(\dx t; q_0)$ (lower right panel) of the Kramers-Moyal expansion
  as defined in (\ref{eq:kramersMoyalDispersion}) for a diffusing
  particle on a half-line with $v=1.0$, $D_q=1.0$ and elapsed time
  $\dx t=0.05$. \cite{noteDimensionless} The solid curves depict the $k_n$
  from the analytical solution
  (\ref{eq:solutionOnedimFokkerPlanck}). The results are obtained
  using one integration step of the Euler scheme. The (red) dashed
  curves correspond to the event-driven method (using
  (\ref{eq:constructProbEvent}) and numerical integration), the (blue)
  dotted line represents the rejection method (with advancing
  time). The event-driven data is only slightly off the exact curve
  (compare insets). The upper graphs on both sides show the probability $w$ for an
  encounter with the wall, see relation (\ref{eq:w}).}
\end{center}
\end{figure}

Due to the presence of the hard wall (reflecting boundary) at the origin,
the statistical properties of the particle displacements
for finite propagation times $\dx t$ deviate from
a Gaussian (compare the Smoluchowski solution (\ref{eq:solutionOnedimFokkerPlanck})).
For a diffusing particle close to a wall, for instance, the first central
moment $\langle q-q_0\rangle$ will not be equal to $v\dx t$ and thus will
not vanish even for absent deterministic drift.
Only in the limit
$\sqrt{2D_q\dx t}/q_0 \to 0$ (and $q_0>0$) the 'Gaussian' statistics is
restored, i.\,e. for small propagation time and/or  sufficiently far away from the
wall, so that its influence is practically not noticeable.
For analyzing how well the different integrations schemes
reproduce these deviations from a Gaussian behavior close to the wall,
we quantify them by looking at the $n$th
expansion coefficient of the Kramers-Moyal series
\cite{Gardiner1983,Risken1984,vanKampen1987} that is associated with
the transition probability (\ref{eq:solutionOnedimFokkerPlanck}) of a
diffusing particle on a half-line subject to a constant force:
\begin{equation}
\label{eq:kramersMoyalDispersion}
  k_n(\dx t; q_0) = \frac{1}{n! \dx t} \int\limits^\infty_{-q_0} \dx \zeta \zeta^n p(q_0+\zeta, \dx t; q_0).
\end{equation}
We remark that relation
(\ref{eq:kramersMoyalDispersion}) is used to estimate the drift term
and the diffusion coefficient in Fokker-Planck or Langevin equations,
respectively, from empirical data. \cite{Friedrich2000,Friedrich2011}
For $q_0>0$ one has $\lim_{\dx t \to 0} k_1(\dx t; q_0) = v$,
$\lim_{\dx t \to 0} k_2(\dx t; q_0) = D_q$ and $\lim_{\dx t \to 0}
k_n(\dx t; q_0) = 0$ for all $n>2$, i.\,e. Gaussian statistics is
recovered in the limit of vanishing time steps $\dx t$.
For finite $\dx t$, the second and
third coefficients are shown in Fig. \ref{bild:dispersionAtWall} as
a function of the initial separation from the wall for the different
integration schemes. The moments exhibit a strong dependence on $q_0$ close
to the wall and show a non-Gaussian behavior. For example, the third
moment 'vanishes' only for considerable distances from the wall. In
general, for increasing distance $q_0$ the coefficients approach the 
Gaussian values
for unrestricted diffusion in a constant force field ('bulk-like'
situation), namely $D_q + \frac{1}{2}v^2 \dx t$ for $k_2$ and $D_qv\dx t
+ \mathcal{O}(\dx t^2)$ for $k_3$.

\begin{figure}[h!]
\begin{center}
\includegraphics[scale=0.325,angle=0]{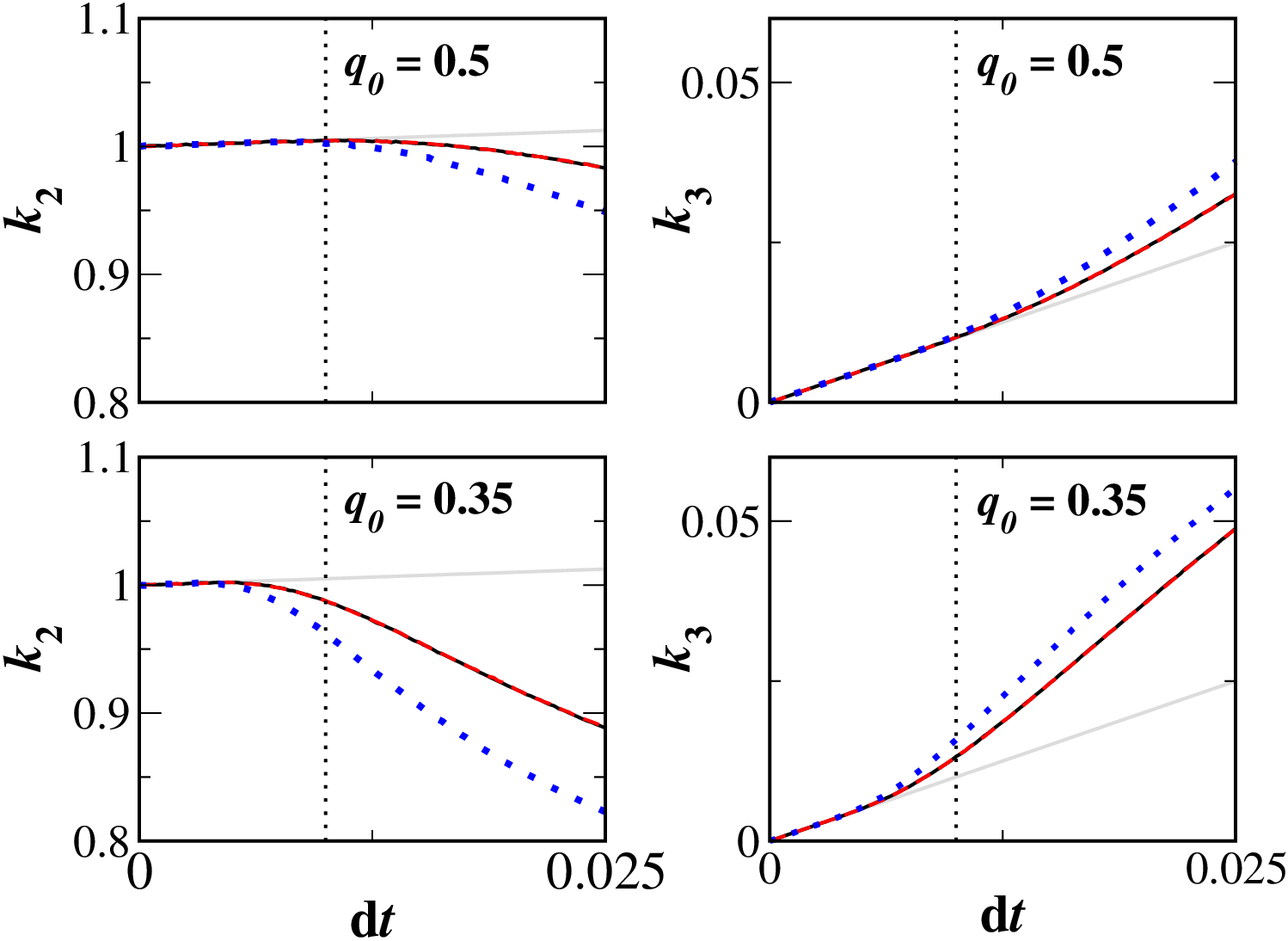}
\caption{\label{bild:dispersionAtWallDT} Second and third
 ('finite-time') coefficient $k_2(\dx t; q_0)$ (left hand side) and
 $k_3(\dx t; q_0)$ (right hand side) of the Kramers-Moyal expansion
 as defined in (\ref{eq:kramersMoyalDispersion}) for the different
 algorithms as a function of the time step $\dx t$ ($v=1.0$, $D_q=1.0$)
 to illustrate their convergence. \cite{noteDimensionless} Suppose
 the particle to be propagated by $\dx t$ is located at
 $q_0=0.5$. The coefficients $k_2$ and $k_3$ are displayed in the
 upper panels. The black curves represent the exact expressions (by
 construction identical to our algorithm), the (red) dashed curves the
 event-driven scheme and the (blue) dotted curves the rejection
 method. (The light grey curves show the asymptotic expressions for
 $\dx t \to 0$, 'bulk-like' situation, see main text). For the choice
 $\dx t=0.01$ (vertical dashed lines) $k_2$ and $k_3$ are well
 reproduced by the rejection scheme.  Suppose now that the particle
 is propagated to $q=0.35$. The coefficients for the next time step
 starting then at $q_0=0.35$
 are displayed in the lower panels. Now, the rejection scheme shows
 considerable deviation in $k_2$ and $k_3$ for the originally adjusted $\dx
 t = 0.01$.}
\end{center}
\end{figure}

From Fig. \ref{bild:dispersionAtWall} we see that in particular
close to the wall
the rejection scheme shows significant deviations from the exact solution.
Its convergence properties do
not only depend on the choice of $\dx t$ with respect to the force
field, but also on how close a
trajectory approaches the wall.  Figure \ref{bild:dispersionAtWallDT} illustrates this
mutual dependence of a proper $\dx t$ and the distance by analyzing
how good the statistical properties of the stochastic evolution are
captured. In contrast, the choice of $\dx t$ for our algorithm is not
influenced by the distance from the wall and is solely limited by the variations in the
force field.

\section{Hard walls}
\label{sec:hardWall}

Our novel procedure to incorporate hard-body
interactions into a Brownian dynamics simulation is now developed for
space-fixed walls. The walls might be curved so that their normal
vectors attain a position dependence.  The discussion will be carried
out for point like Brownian particles. For extended spherical
particles the distance to the hard surface is reduced by the radius of
the particle.

To develop the method independently of the used algorithm for the
numerical integration we discuss the procedure on the level of the
Langevin equation (\ref{eq:langevinWithWiener}) formulated in terms of
differentials with the tacit understanding that $\dx \vektor{r}$ has
to be seen as the numerically obtained displacement for the (discrete)
time step $\dx t$.  We also assume that the time step is already
chosen appropriately small to account for variations in the force
fields.

\subsection{Description of algorithm}

Suppose that the Brownian particle is
currently at position $\vektor{r}$ close to the wall and experiences
the deterministic force $\vektor{F}$.
Then, the standard integration scheme, which 'suggests' a displacement $\dx\vektor{r}$,
is modified in the following way to yield a physically valid displacement
$\dx\vektor{r}^\stern$, which accounts for the hard wall:
\begin{enumerate}
\item[1.]
Calculate a 'preliminary' displacement $\dx \vektor{r}$
using the standard integrator for the 'free' Langevin equation without wall.
\item[2.]
Check whether or not the so suggested new position $\vektor{r}+\dx\vektor{r}$
yields a physically valid configuration (i.\,e. whether or not the particle, if
at position $\vektor{r}+\dx\vektor{r}$, overlaps with the hard wall).
\begin{enumerate}
\item[a.]
If this configuration is valid, the suggested displacement is accepted
as the new one, $\dx \vektor{r}^\stern = \dx \vektor{r}$. 
\item[b.] 
If this configuration is invalid, the new, valid displacement is calculated
according to
\[
\dx \vektor{r}^\stern = \dx \vektor{r} + (q-q_0 - \vektor{n}\cdot \dx \vektor{r}) \vektor{n} \, ,
\]
where $\vektor{n}$ is the normal on the wall that
passes through $\vektor{r}$,
$q_0$ is the distance of the particle at initial position
$\vektor{r}$ from the wall, and $q$ is generated from
$p_\indexHB(q,q_0,\dx t,v, D_q) $ with $v=\vektor{n}\cdot\vektor{F}/\eta$
and $D_q = D$ (compare relation (\ref{eq:distributionOutwardWall}) and Appendix \ref{sec:generateRandomQ}).
\end{enumerate}
\end{enumerate}
We note that $\dx t$ has to be adjusted such that $\lambda \gg
\mathcal{O}(|\dx \vektor{r}|)$ holds true with $\lambda$ being the
smallest radius of curvature of the wall.

\subsection{Justification and discussion}

For the time step that has
produced the unphysical configuration the displacement can be decomposed into
a displacement perpendicular to the wall (along the normal $\vektor{n}$),
\begin{eqnarray}
  \dx \vektor{r}_\perp = (\vektor{n}\cdot \dx \vektor {r}) \vektor{n}
  = \frac{\vektor{F}_\perp}{\eta} \dx t + \sqrt{2D} \dx
  \vektor{W}_\perp \, ,
\end{eqnarray}
and a displacement parallel to the wall,
\begin{eqnarray}
  \dx \vektor{r}_\parallel = \dx \vektor{r} - (\vektor{n}\cdot \dx \vektor {r})\vektor{n} = \frac{\vektor{F}_\parallel}{\eta} \dx t+ \sqrt{2D} \dx \vektor{W}_\parallel \, .
\end{eqnarray}
The perpendicular component of the deterministic
force is given by $\vektor{F}_\perp = (\vektor{n}\cdot\vektor{F})
\vektor{n}$ whereas the parallel component is $\vektor{F}_\parallel =
\vektor{F} - \vektor{F}_\perp$. The components of the fluctuating
force are obtained correspondingly and are given by $\dx
\vektor{W}_{\perp} = (\dx \vektor{W}\cdot \vektor{n})\vektor{n}$ and $
\dx \vektor{W}_{\parallel} = \dx \vektor{W} - (\dx \vektor{W}\cdot
\vektor{n})\vektor{n}$.  These two components are uncorrelated due to
the orthonormality of the perpendicular and parallel directions:
\begin{eqnarray}
\langle \dx W^\alpha_{\perp}\dx W_{\parallel}^\beta\rangle 
&=& \langle \dx W^\mu n^\mu n^\alpha( \dx W^\beta - \dx W^\nu n^\nu n^\beta)\rangle \\
&=& \dx t  n^\alpha n^\beta(1- n^\mu n^\mu) = 0.
\end{eqnarray}
The correlations need to be considered only for the current time step
where an unphysical position is generated. The normal
$\vektor{n}$ sets up a locally fixed coordinate system.
However, one is not interested in its dynamical properties so that its (possible)
position dependence has not to be considered in a dynamical
(co-moving) sense. For this particular time step the random forces in
the two equations are those that appear at time $t$, for the
corresponding diffusion constants one has $D_\perp = D_\parallel =
D$. It is essential that the perpendicular and parallel components
of the displacement decouple and decorrelate
for this local coordinate, so that they can be considered
independently.

Let us assume for the moment that we have an appropriately small time
step so that the typical displacement $\dx \vektor{r}$ is small
compared to the (smallest) radius of curvature of the point on the
wall that corresponds to the normal $\vektor{n}$.\cite{noteMultipleCollisions}
Hence, the wall
appears to be flat on the scale of $\dx \vektor{r}$. 
In this approximation, the perpendicular
motion along the normal $\vektor{n}$ thus corresponds to a motion on
an infinite half-line, whereas the motion along
the parallel direction is unrestricted. The overlap with the wall then
solely results from the (one-dimensional) displacement along
$\vektor{n}$. A valid physical end position along that direction, i.\,e. the
distance $q$ from the wall after
elapsed time $\dx t$, is now directly generated from
(\ref{eq:distributionOutwardWall}), with $q_0$ being the separation of
the initial position $\vektor{r}$ from the wall, $D_q = D$ and $v =
\vektor{n}\cdot \vektor{F} / \eta$.

The parallel motion is uncorrelated with the perpendicular motion for
the considered time step on the level of the fluctuating forces and
therefore is unaffected by that modification. We note that the
deterministic force is fixed for the considered time step and that the
modification accounts for the altered statistical properties of the
random forces in the vicinity of a hard wall. Hence, the already
proposed displacement $\dx \vektor{r}_\parallel$ can be retained to
avoid unnecessary computational efforts. The total modified physical
displacement is then given by $\dx \vektor{r}^\stern = \dx q
\vektor{n} + \dx \vektor{r}_\parallel$ with $\dx q= q - q_0$. Plugging
in the originally proposed $\dx \vektor{r}$ on gets the modified $\dx
\vektor{r}^\stern = \dx \vektor{r} + (\dx q - \vektor{n}\cdot \dx
\vektor{r}) \vektor{n}$ already given above.

We note that the maximum allowable size of the integration time step
$\dx t$ is limited by the curvature of the hard wall. However, the
length scales of physical systems for which Brownian dynamics is
applied is typically mesoscopic so that walls are only weakly curved
on the scale of typical displacements and thus the corresponding
radius of curvature is large. To be more explicit, consider
protein-protein association as a paradigmatic example. Here the
extension of (space-fixed) obstacles is usually comparable to (or
even larger than) the diameter of the proteins themselves, say
40\AA. The range of the involved non-covalent forces is typically of
the order of 2 to 4\AA. \cite{Leckband2001, Jackson2006,Schreiber2009}
Therefore, the time step that has to be used to capture the variations
in the force fields is already small enough to account for the
curvature of the obstacles as well. In summary, the local flatness
of the boundary is necessary for our algorithm to be
applicable, but this approximation
can be controlled by the choice of the size of the time step
very much in the same way as space-dependent variations of the forces
are captured by sufficiently small time steps.

Finally, we want to point out that in the force-free case $v=0$, the
solution (\ref{eq:solutionOnedimFokkerPlanck}) simplifies to
$p(q,t;q_0)=p_1(q,t;q_0)+p_1(q,t;-q_0)$ (since
$p_2(q,t;q_0)=p_1(q,t;-q_0)$ and $p_3(q,t;q_0)=0$, compare
(\ref{eq:p1dimForm})), consistent with what one would obtain from the
so-called image method (see also Appendix \ref{sec:OU}). As a
consequence, our new algorithm, which is constructed from expressions
(\ref{eq:solutionOnedimFokkerPlanck}) and (\ref{eq:p1dimForm}),
becomes equivalent (in a strict mathematical sense) to numerical
schemes which replace invalid positions by an image end-point whenever
a wall is encountered and externally applied forces are absent. A
prominent example is the algorithm developed by Scala et
al. \cite{Scala2007}

\section{Hard spheres}
\label{sec:hardSpheres}

In the previous section physical systems with space-fixed hard walls
have been considered. In many systems, however, movable objects with
hard cores are present as well. In this section we describe, how the
previously discussed method can be extended to encounters of two hard
spheres, which both move under the influence of deterministic forces
and diffusion.

\subsection{Description of algorithm}

Consider two hard spheres with
(possibly different) radii $a_i$, friction coefficients $\eta_i$ and
diffusion constants $D_i=\kb T/\eta_i$ whose dynamics is described by the two
Langevin equations
\begin{subequations}
\label{eq:twoSpheresLangevin}
\begin{eqnarray}
\label{eq:twoSpheresLangevin1}
\dx{\vektor{r}}_1 &=& \vektor{v}_1 \dx t + \sqrt{2D_1}\dx \vektor{W}_1
\end{eqnarray}
and
\begin{eqnarray}
\label{eq:twoSpheresLangevin2}
\dx{\vektor{r}}_2 &=& \vektor{v}_2 \dx t + \sqrt{2D_2}\dx \vektor{W}_2.
\end{eqnarray}  
\end{subequations}
Here again the drift terms are denoted in terms of velocities
$\vektor{v}_i = \vektor{F}_i(\vektor{r}_i,t)/\eta_i$ for the two particles.

Suppose now that the two spheres are at $\vektor{r}_1$ and
$\vektor{r}_2$ at time $t$ and that we have at hand a standard
integration scheme to update the particle positions for
a time step $\dx t$. Then, the algorithm for generating
valid displacements $\dx \vektor{r}_1^\stern$ and
$\dx \vektor{r}_2^\stern$ consistent with
hard-wall interactions between these two particles
is as follows:
\begin{enumerate}
\item[1.]
Use the standard integration scheme to generate
'preliminary' propagation steps $\dx \vektor{r}_1$, $\dx \vektor{r}_2$.
\item[2.]
Check whether or not the suggested end positions $\vektor{r}_1+\dx\vektor{r}_1$
and $\vektor{r}_2+\dx\vektor{r}_2$ represent a physically valid configuration, i.\,e. whether or not the two spheres at positions $\vektor{r}_1+\dx\vektor{r}_1$
and $\vektor{r}_2+\dx\vektor{r}_2$ would overlap.
\begin{enumerate}
\item[a.]
If the two spheres do not overlap, the new configuration is accepted,
$\dx\vektor{r}_1^\stern=\dx\vektor{r}_1$ and $\dx\vektor{r}_1^\stern=\dx\vektor{r}_1$.
\item[b.]
If the two spheres do overlap,
valid displacements $\dx\vektor{r}_1^\stern$, $\dx\vektor{r}_2^\stern$
are constructed according to the following procedure:
\begin{itemize}
\item Calculate the separation $q_0 = |\vektor{r}_2-\vektor{r}_1|-(a_1+a_2)$
  of the surfaces of the two spheres and determine the
  normalized connection vector
  $\vektor{e}=(\vektor{r}_2-\vektor{r}_1) / |\vektor{r}_2-\vektor{r}_1|$.
\item Generate a position $q$ from $p_\indexHB(q,q_0,\dx t,v, D_q)$
  with diffusion constant $D_q = D_1 + D_2$ and velocity $v=(\vektor{v}_2 - \vektor{v}_1)\cdot \vektor{e}$
  (compare relation (\ref{eq:distributionOutwardWall}) and Appendix \ref{sec:generateRandomQ}).
\item Use $\dx q = q - q_0$ to evaluate the modified physical displacements 
\begin{subequations}
\label{eq:drTwoSpheresStern}
\begin{equation}
\label{eq:dr1stern}
\hspace*{2em} \dx \vektor{r}_1^\stern = \dx \vektor{r}_1 + \frac{\eta_2}{\eta_1+\eta_2}\left( (\dx\vektor{r}_2 - \dx\vektor{r}_1)\cdot\vektor{e} - \dx q \right)\vektor{e} 
\end{equation}
 and 
\begin{equation}
\label{eq:dr2stern}
\hspace*{2em} \dx \vektor{r}_2^\stern = \dx \vektor{r}_2 - \frac{\eta_1}{\eta_1+\eta_2}\left( (\dx\vektor{r}_2 - \dx\vektor{r}_1)\cdot\vektor{e} - \dx q \right)\vektor{e} \, .
\end{equation}
\end{subequations}
\end{itemize}
\end{enumerate}
\end{enumerate}
The time step $\dx t$ has to by chosen small enough so that $a_1+a_2
\gg \mathcal{O}(|\dx \vektor{r}_2 - \dx \vektor{r}_1|)$ is satisfied.

\subsection{Justification and discussion}

To generate a physical configuration without
overlap the motion of the two Brownian particles is decomposed into a
common center-of-friction motion and a relative motion of the two
spheres. 
The center-of-friction $\vektor{R} =
(\eta_1 \vektor{r}_1 + \eta_2\vektor{r}_2)/\eta$, $\eta = \eta_1 +
\eta_2$,
obeys the Langevin equation
\begin{eqnarray} 
  \dx {\vektor{R}} &=& \frac{1}{\eta}\left( \eta_1 \vektor{v}_1 +
  \eta_2\vektor{v}_2\right) \dx t \nonumber \\ && + \frac{1}{\eta}
  \left( \eta_1\sqrt{2D_1}\dx \vektor{W}_1 + \eta_2\sqrt{2D_2} \dx
  \vektor{W}_2 \right) \\
\label{eq:centreOfFrictionMotion}
& =& \vektor{V}\dx t + \sqrt{2D_R} \dx \vektor{W}_R
\end{eqnarray}
and the relative motion of $\vektor{r} = \vektor{r}_2 - \vektor{r}_1$ is governed by
\begin{eqnarray}
  \dx {\vektor{r}} &=& (\vektor{v}_2 - \vektor{v}_1) \dx t + \sqrt{2D_2}\dx \vektor{W}_2 - \sqrt{2D_1}\dx \vektor{W}_1\\
\label{eq:relativeMotion} & =& \vektor{v}\dx t + \sqrt{2D_r} \dx \vektor{W}_r.
\end{eqnarray}
Here, the abbreviations $ \vektor{v} := \vektor{v}_2 - \vektor{v}_1$
and $\vektor{V} := \frac{1}{\eta}\left( \eta_1 \vektor{v}_1 +
\eta_2\vektor{v}_2\right) = \frac{1}{\eta}( \vektor{F}_1 +
\vektor{F}_2)$ have been introduced.  The fluctuation terms in
(\ref{eq:centreOfFrictionMotion}) and (\ref{eq:relativeMotion}) can
again be written in terms of un-correlated white noise terms $\dx
\vektor{W}_R$ and $\dx \vektor{W}_r$ with associated diffusion
coefficients $D_R = \kb T /\eta = D_1D_2/(D_1+D_2)$ and $D_r = D_1 +
D_2$, respectively.  The statistical properties of the fluctuation
terms in (\ref{eq:centreOfFrictionMotion}) and
(\ref{eq:relativeMotion}) are readily obtained from the observation
that the sum and the difference of two Wiener processes are again
Wiener processes (as theses processes are Gaussian stochastic
processes with vanishing mean). Due to the appropriate weighting of
the contributions of the two particles in the definition of the
center-of-friction, $\dx \vektor{W}_R$ and $\dx \vektor{W}_r$ are
uncorrelated, as can be seen directly by equating the cross
correlations.

As a consequence, any corrections in the relative motion do
not affect the propagation in center-of-friction $\vektor{R}$ (and vice versa).
The overlap of the two particles is the result of the unphysical
propagation in the relative coordinate $\vektor{r}$ so that this
very motion has to be modified due to the hard-wall interaction,
while the initially proposed displacement $\dx \vektor{R}$ of the
center-of-friction can be kept
to avoid unnecessary computational efforts.
This is similar to the situation
of a particle in the vicinity of a wall where the normal and the
parallel displacements decorrelate.

The relative motion itself corresponds to the motion of a
point-particle close to a space-fixed sphere of radius $a_1 + a_2$.
The treatment of the relative motion therefore follows the lines of
reasoning already developed in Sec. \ref{sec:hardWall}. The
relative displacement $\dx \vektor{r}$ can be further decomposed into
a (one-dimensional) displacement $\dx \vektor{r}_\perp$ along the
connection line defined by the unit vector $\vektor{e} :=
\vektor{r}/|\vektor{r}|$ and a displacement $\dx \vektor{r}_\parallel$
in the plane perpendicular to $\vektor{e}$.  Thus one obtains the
equations
\begin{subequations}
\begin{equation}
\label{eq:RelDrPerp}
\dx r_\perp= \vektor{v}\cdot\vektor{e} \,\dx t + \sqrt{2D_r} \,\dx \vektor{W}_r\cdot \vektor{e} 
\end{equation}
and 
\begin{equation}
\label{eq:RelDrParallel}
\dx \vektor{r}_\parallel= (\vektor{v} -\vektor{v}_\perp) \dx t +
\sqrt{2D_r} \left(\dx \vektor{W}_r - (\dx \vektor{W}_r\cdot
\vektor{e})\vektor{e} \right)
\end{equation}
\end{subequations}
with $\vektor{v}_\perp = (\vektor{e}\cdot\vektor{v})\vektor{e}$. (Note
that $\vektor{e}$ is the normal of the effective hard surface that
restricts the motion of the relative coordinate. The nomenclature of
perpendicular and parallel displacement has to be seen with respect to
this spherical surface). The crucial point is that the perpendicular
fluctuating term $ \dx {W}_{\perp} = \dx \vektor{W}_r\cdot \vektor{e}$
and the parallel random term $ \dx \vektor{W}_{\parallel} = \dx
\vektor{W}_r - (\dx \vektor{W}_r\cdot \vektor{e})\vektor{e}$ are again
un-correlated (for the current time step where the overlap is
detected).  The associated diffusion constants are $D_\perp =
D_\parallel = D_r$.  The vector $\vektor{e}$ is fixed for the
considered time step and again defines a local coordinate
system suitable for the decomposition.

Assuming that the effective hard wall (that corresponds to a sphere of radius
$a_1+a_2$) can be well approximated as flat
(which is, for example, well fulfilled in the typical case in point
of protein-protein association, see Sec.
\ref{sec:hardWall}),
the relative motion along the direction
$\vektor{e}$ then corresponds to the motion on an infinite half-line
and the parallel motion is unrestricted in an infinite
plane. Therefore, the described algorithm of Sec.
\ref{sec:Brownian} for the one-dimensional system can be applied
directly to generate a new physical position along $\vektor{e}$.  The
propagation in the plane perpendicular to $\vektor{e}$ is not affected
by that modification and thus the corresponding
initially obtained displacement can be kept.  To satisfy the
assumption that the hard surface is virtually flat the time step has
to be chosen sufficiently small so that one has $a_1+a_2 \gg
\mathcal{O}(|\dx \vektor{r}|)$.\cite{noteMultipleCollisions}

The new configuration without overlap of the spheres is obtained from
the modified relative displacement by
\begin{equation}
\label{eq:newDisplacementSpheres}
  \dx \vektor{r}^\stern = \dx \vektor{r}_\parallel  + \dx q \vektor{e}
\end{equation}  
with $\dx \vektor{r}_\parallel = \dx \vektor{r} - (\dx
\vektor{r}\cdot\vektor{e})\vektor{e}$ being the originally proposed
parallel displacement ($\dx \vektor{r}$ denotes the initially proposed
relative displacement $\dx \vektor{r}_2 - \dx \vektor{r}_1$ of the two
spheres). Here, $\dx q = q -q_0$ is fixed by $q$ that is generated
from distribution (\ref{eq:distributionOutwardWall}) with diffusion
constant $D_q=D_\perp=D_1+D_2$, velocity $v =
\vektor{v}\cdot\vektor{e} = (\vektor{v}_2 - \vektor{v}_1)\cdot
\vektor{e}$, initial separation $q_0 = |\vektor{r}_2-\vektor{r}_1| -
(a_1+a_2)$ of the two spheres.  Once $\dx \vektor{r}^\stern$ is
determined the new absolute particle positions without overlap are
calculated from the new displacements as
\begin{subequations}
\label{eq:finalDisplacementsSpheres}
\begin{eqnarray}
\dx \vektor{r}_1^\stern &=& \dx \vektor{R} - \frac{\eta_2}{\eta_1 + \eta_2} \dx \vektor{r}^\stern
\end{eqnarray}
and
\begin{eqnarray}
\dx \vektor{r}_2^\stern &=& \dx \vektor{R} +\frac{\eta_1}{\eta_1+\eta_2} \dx \vektor{r}^\stern.
\end{eqnarray} 
\end{subequations}
Plugging in the explicit expressions for the various quantities one
ends up with equations (\ref{eq:dr1stern}) and (\ref{eq:dr2stern}).

\section{Exemplary model systems}
\label{sec:exemplaryModels}

In this section we numerically investigate the applicability and
quality of the new algorithm in comparison with the heuristic methods
of Sec. \ref{sec:hardWallHeuristics}. We do not aim at a study of
particular physical systems or phenomena, we rather intend to present
exemplary problems that typically have to be dealt with in systems
modelled within Brownian dynamics. To set up an explicit frame we look
at two research areas from soft matter, namely
the motion of colloids in flow fields and the interaction of
proteins in solution. For
simplicity we consider only two-dimensional systems. All numerical
results presented below are obtained by using the Euler algorithm
as standard integrator.

\subsection{Colloids in flow fields}

The first characteristic system we
look at is given by colloidal particles in 
stationary flow fields $\vektor{u}(\vektor{r})$. 
To begin with we consider 
a spherical particle of radius $a=1\,\mu$m and with diffusion
constant $D=0.21\,\mu$m$^2$/s
(Stokes-Einstein diffusion at room temperature $\kbT \approx 4\,\mbox{fN}\mu\mbox{m}$)
suspended in the fluid flow
around a space-fixed circular post in two space dimensions.  This
model system is taken to mimic a typical situation in microfluidic
devices where particles encounter posts or walls in the
channel.  Let us consider a flow field with a typical velocity of $u_0
= $10$\,\mu$m/s (typical velocities in microfluidic channels range
from 1$\,\mu$m/s to 1\,cm/s) around a circular post of radius
$R=10\mu$m at the origin of the coordinate system. For a time step
$\dx t = 0.01$\,s the typical diffusive displacement
$\sqrt{\langle\vektor{r}^{\,2}\rangle-\langle\vektor{r}\rangle^2}$
of a particle with radius $a=1\,\mu$m is
0.1\,$\mu$m. Thus we deduce that for a radius of $R = 10\,\mu$m of the
post the curvature of the hard surface should not limit the
applicability of our method. Assuming no-slip boundary conditions, the
velocity field of the two-dimensional stationary Stokes
flow $\vektor{u}(\vektor{r}) = (u_x(x,y), u_y(x,y))$ around the post
is given by
\begin{subequations}
\label{eq:flowPastPost}
\begin{equation}
u_x = u_0 \frac{y^2-x^2}{x^2+y^2} \left( 1- \frac{R^2}{x^2+y^2}\right) + 2u_0\ln\frac{\sqrt{x^2+y^2}}{R} 
\end{equation}
and
\begin{equation}
u_y = - u_0\frac{xy}{x^2+y^2} \left( 1- \frac{R^2}{x^2+y^2}\right)
\end{equation}
\end{subequations}
in Cartesian coordinates (for a discussion of the Stokes approximation
for small Reynolds numbers and its validity in two space dimensions, see for example
Refs. \onlinecite{Proudman1957,Acheson1990}). 

\begin{figure}[h!]
\begin{center}
\includegraphics[scale=0.355,angle=0]{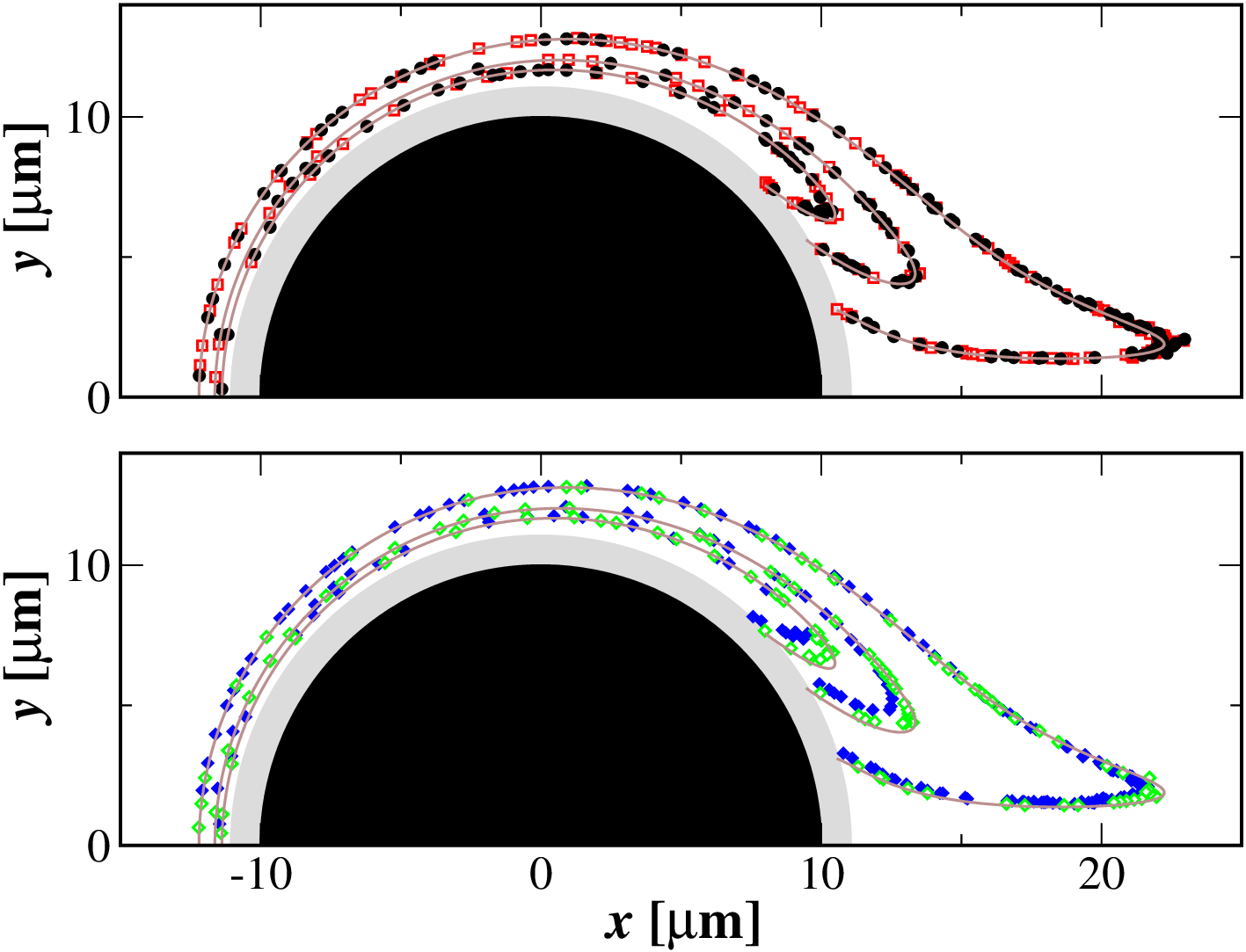}
\caption{\label{bild:flowAroundSphere} Contour plot of the
  distribution of a spherical colloid (radius $a=1\,\mu$m, diffusion
  constant $D=0.21\,\mu{m}^2$/s) flowing around a circular obstacle of
  radius $R=10\,\mu$m. The particle starts initially at $\vektor{r}(0)
  = (-11.5\,\mu\mbox{m}, 0\,\mu\mbox{m})$ and evolves for $t = 10\,$s.
  The Stokesian flow (\ref{eq:flowPastPost}) is characterized by
  $u_0 = 10\,\mu$m/s. The solid line represents the solution of the
  associated Fokker-Planck equation. In the upper panel symbols
  display the numerical solutions obtained from integrating the
  Langevin equation with $\vektor{v}(\vektor{r})=\vektor{u}(\vektor{r})$
  for a time step $\dx t = 0.01\,$s
  with circles corresponding to our algorithm and open (red)
  squares to the event-driven scheme. The lower panel shows the data
  from the rejection scheme for $\dx t = 0.01\,$s (blue diamonds) and
  $\dx t = 0.001\,$s (open green diamonds).
  The black region represents the obstacle,
  the grey shaded region is inaccessible to the particle coordinate.}
\end{center}
\end{figure}

\begin{figure}[h!]
\begin{center}
\includegraphics[scale=0.35,angle=0]{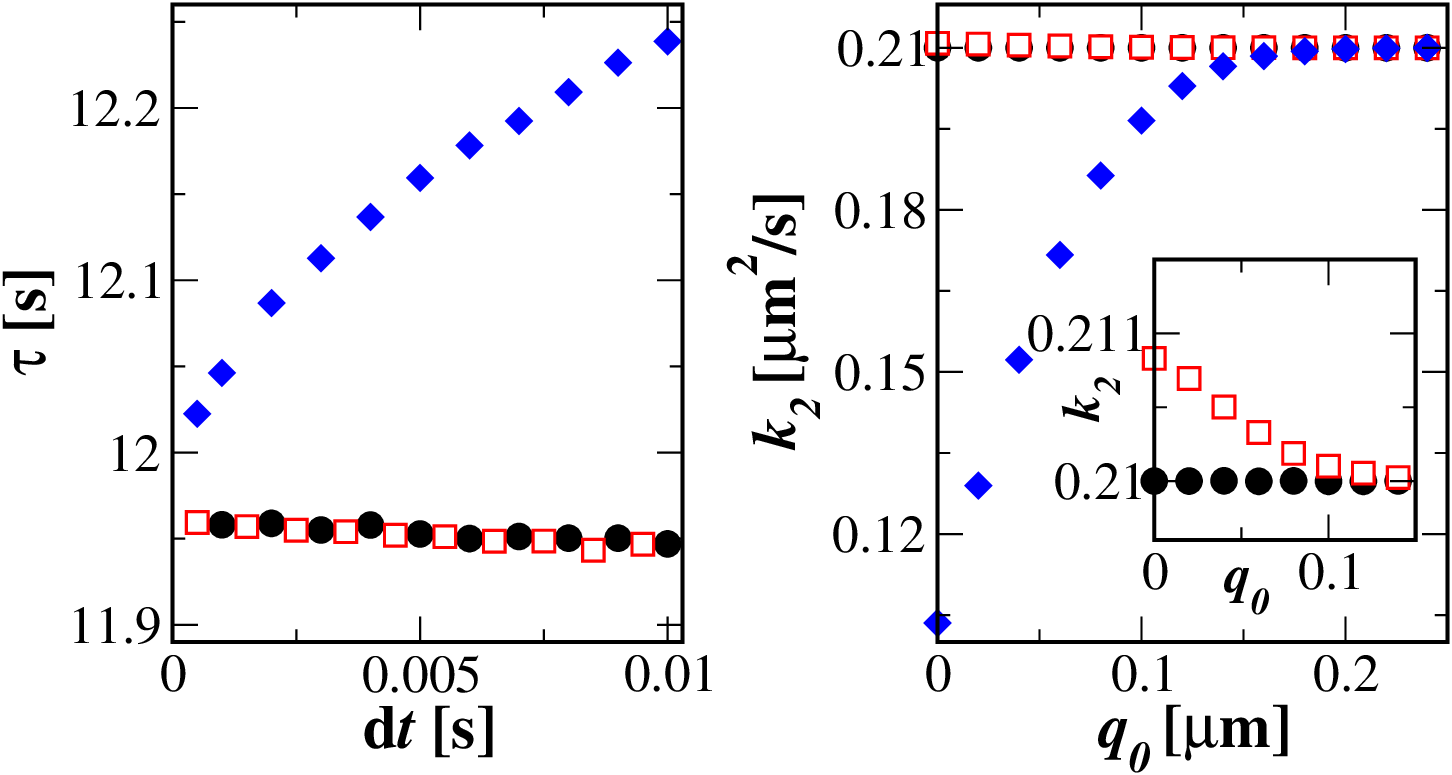}
\caption{\label{bild:passageTime} Left panel: Mean first passage time
  $\tau$ a colloid of radius $a=1\,\mu$m needs to flow past a post of
  radius $R=10\,\mu$m to the threshold at $x = b=11.5\,\mu$m as a function
  of the integration time step. The colloid starts initially at
  $\vektor{r}(0) = (-11.5\,\mu\mbox{m}, 0\,\mu\mbox{m})$. Our method is
  displayed by (black) circles, the event-driven scheme by (red) open
  squares and the rejection scheme by (blue) diamonds.  For the
  numerical determination of the mean first passage time, see e.\,g.
  Refs. \onlinecite{Honerkamp1990,Mannella1999}. Note that the
  collision frequency with the post drops from 3\% to 1\% for the
  shown range of $\dx t$. Right panel and inset: Tangential diffusion
  $k_2 = \langle \left(\dx \vektor{r}\cdot
  \vektor{e}_\parallel\right)^2\rangle/(2\dx t)$ during the first time
  step of size $\dx t = 0.01$\,s as a function of the initial distance
  $q_0 = |\vektor{r}(0) - (R + a)|$ between the surfaces of the
  colloid and the post. The different schemes are displayed using the
  symbol code of the left panel. The expected value is
  $0.21\,\mu\mbox{m}^2$/s.}
\end{center}
\end{figure}

Figure \ref{bild:flowAroundSphere} compares the numerically determined
distribution function of a colloid in flow field
(\ref{eq:flowPastPost}) with the solution of the associated
Fokker-Planck equation. The evolution of the particle as exhibited by
the rejection scheme lags behind the expected behavior. To analyze
this further we look at the mean first passage time the colloid needs
to cross a certain threshold after passing the post. Figure
\ref{bild:passageTime} displays the corresponding results as a
function of the integration step $\dx t$.
It shows also the particle diffusion $k_2 = \langle \left(\dx
\vektor{r}\cdot \vektor{e_\parallel}\right)^2\rangle/(2\dx t)$
tangential to the post during the first time step as a function of the
initial separation from the post's surface ($\vektor{e}_\parallel$
denotes the tangential unit vector to the surface). The tangential
motion is not restricted by the presence of the hard surface so that
$k_2$ is expected to be given by the diffusion constant $D$ of the
particle. For the rejection scheme, however, the tangential diffusion
close to the wall is clearly reduced. This has the effect that the
time the particle needs to flow around the obstacle is increased when
measured from the rejection scheme. As can be seen in Fig.
\ref{bild:passageTime} this effect is reduced for decreasing time
steps as the collision frequency with the obstacle goes down as well.

\begin{figure}[h!]
\begin{center}
\includegraphics[scale=0.355,angle=0]{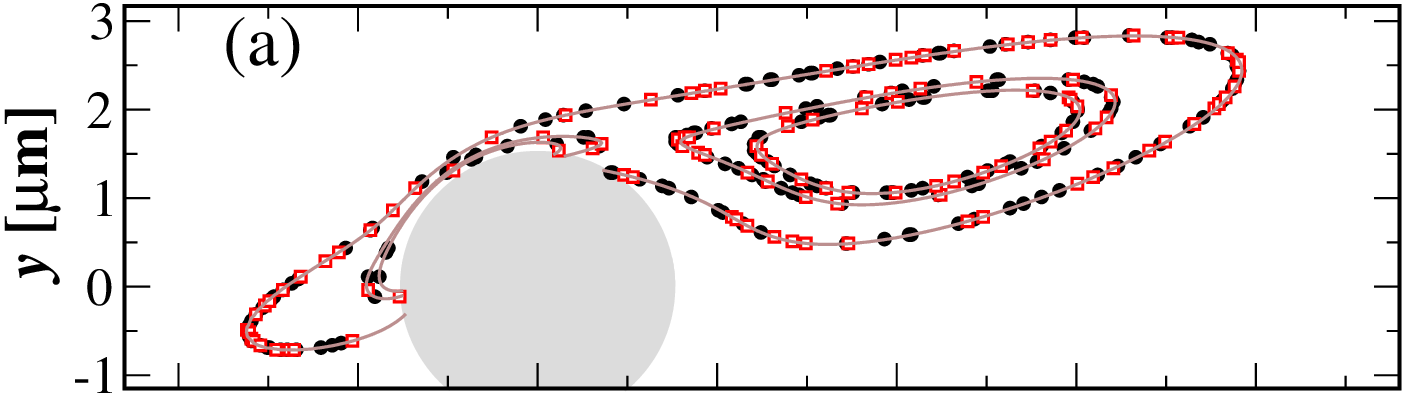}

\includegraphics[scale=0.355,angle=0]{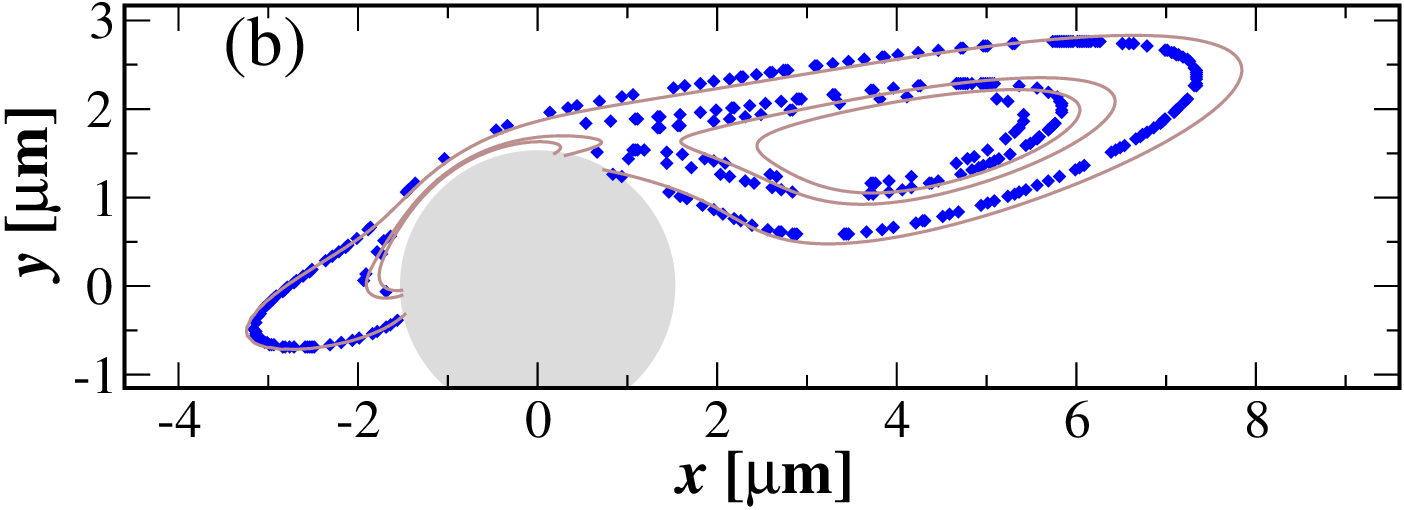}

\includegraphics[scale=0.355,angle=0]{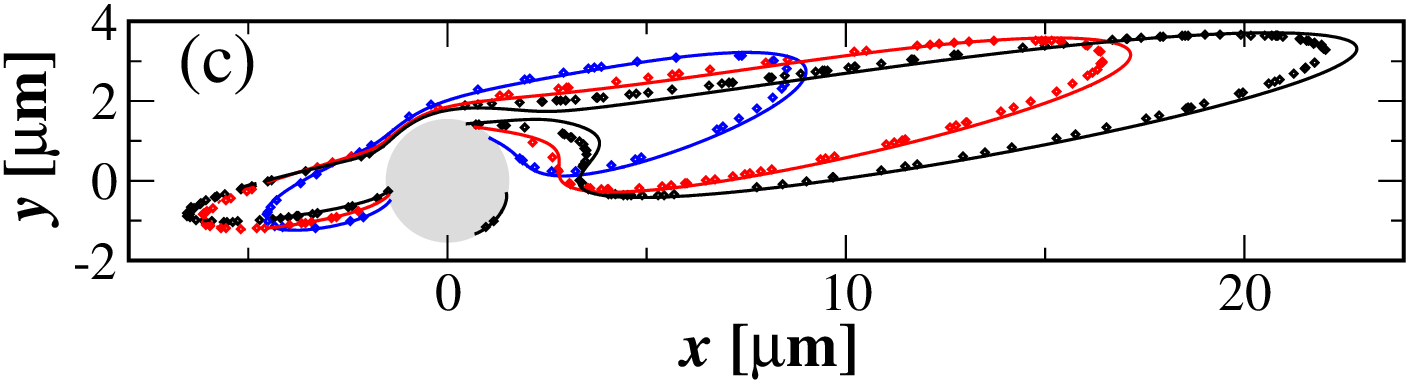}
\caption{\label{bild:shearFlow} Contour plot of the transition
  probability of the relative vector $\vektor{r} = (x,y)= \vektor{r}_2
  - \vektor{r}_1$ of two colloids in a shear flow (\ref{eq:shearFlow})
  with shear rate $\alpha = 10$/s. The two particles with radii
  $a_1=1\,\mu$m, $a_2=0.5\,\mu$m and friction coefficients $\eta_1 =
  18.9$ fNs/$\mu$m, $\eta_2 = 9.45$ fNs/$\mu$m (i.\,e. $D_1 =
  0.21\,\mu\mbox{m}^2$/s, $D_2 = 0.42\,\mu\mbox{m}^2$/s at room
  temperature) initially start at $\vektor{r}_1(0) = (0\,\mu\mbox{m},
  0\,\mu\mbox{m})$ and $\vektor{r}_2(0) = (-2\,\mu\mbox{m},
  0.75\,\mu\mbox{m})$, and evolve for $t=0.5$ s (panels (a) and
  (b)). The solution of the associated Fokker-Planck equations are
  depicted by solid lines, the numerical results from integrating the
  Langevin equations with $\dx t = 0.002\,$s are shown by symbols:
  (Black) circles represent our algorithm, open (red) squares
  correspond to the event-driven scheme (panel (a)), and (blue)
  diamonds to the rejection approach (panel (b)). Panel (c) shows the
  evolution of one contour line with time (0.5, 0.8, 1.0 s) in
  comparison to the results of the rejection scheme. The grey circle
  represents the region which is inaccessible to the relative
  coordinate of the two colloids.}
\end{center}
\end{figure}

As a second typical case study we investigate the evolution of two
colloidal particles in a prescribed shear flow, The velocity field at
$\vektor{r} = (x,y)$ is given by
\begin{equation}
\label{eq:shearFlow}
\vektor{u}(\vektor{r}) = \alpha y \vektor{e}_x
\end{equation}
where $\alpha$ is the shear rate and $\vektor{e}_x$ denotes the unit
vector in $x$-direction. Figure \ref{bild:shearFlow} depicts the
distribution function of the relative coordinate of the two colloids
for different elapsed times. Again numerical data is compared to the
solution of the Fokker-Planck equation.

\subsection{Protein-protein interaction}

A (fairly small) protein has a typical radius of approximately
$a=2$\,nm leading to a diffusion constant
$D=0.1\,\mbox{nm}^2/\mbox{ns}$ at room temperature $\kbT \approx
4\,\mbox{pN}\mbox{nm}$.  Therefore, the typical time a diffusing
protein needs to cross its size is of the order of 10\,ns. The forces
between two proteins are governed by non-covalent interactions which
have a typical energy scale of 2 to 4\,$\kb T_{\scriptsize\mb{room}}$
and a range of 0.2 to 0.4\,nm. Therefore, the velocity in the Langevin
equation (\ref{eq:langevinWithWiener}) is of the order of
5\,$\mbox{nm}/\mbox{ns}$. A time step of $\dx t = 0.001$\,ns has a
typical diffusive displacement $\sqrt{\langle
  \vektor{r}^\power\rangle-\langle \vektor{r}\rangle^2}$ of
approximately 0.02\,nm and should be small enough to capture typical
variations in the force fields ($\mathcal{O}(|\vektor{v}\dx t|)\sim
0.005\mbox{\,nm}$) and to resolve the details of trajectories.

To have an explicit example at hand we will look at the interaction of
proteins such as lysozyme, which has a radius of 1.8\,nm and a
diffusion constant of $0.12\,\mbox{nm}^2/\mbox{ns}$ in aqueous
solution. The force between two proteins stems from repulsive
(screened) electrostatic and from attractive short-ranged van-der
Waals interactions.  Modelling the lysozyme solution as a colloidal
system \cite{Tardieu1999,Pellicane2004} within the
Derjaguin-Landau-Verwey-Overbeek theory the electrostatic interaction
between two proteins of separation $r=|\vektor{r}_2 - \vektor{r}_1|$
is represented by the Debye-H\"uckel contribution
\begin{equation}
\phi_\indexDH(r) =  \frac{K}{r}\exp\left( - \kappa(r-\sigma)\right)
\end{equation}
and the attractive van-der Waals part by
\begin{equation}
\phi_\indexHA(r) = -\frac{A}{12}\left(\frac{\sigma^2}{r^2} + \frac{\sigma^2}{r^2-\sigma^2} + 2 \ln\left(\frac{r^2-\sigma^2}{r^2}\right)\right).
\end{equation}
The total interaction potential is then given by 
\begin{equation}
\label{eq:colloidPotential}
\phi(r) =  \left\{ \begin{array}{ll}
         \phi_\indexDH(r) +\phi_\indexHA(r), & \mbox{if $r \geq \sigma + \delta$}\\
        \infty, & \mbox{if $r < \sigma + \delta$}\end{array} \right. 
\end{equation}
from which one obtains the force $\vektor{F}_i = -\nabla_i \phi(r)$ on
particle $i=1,2$. Here $\kappa$ denotes the screening length, $\sigma$
is the radius of the effective particle (i.\,e.\,the sum of the two radii) and $\delta$ is the Stern layer thickness so that the excluded
hard-core radius is given by $R_\indexHC = \sigma + \delta$.

\begin{figure}[h!]
\begin{center}
\includegraphics[scale=0.315,angle=0]{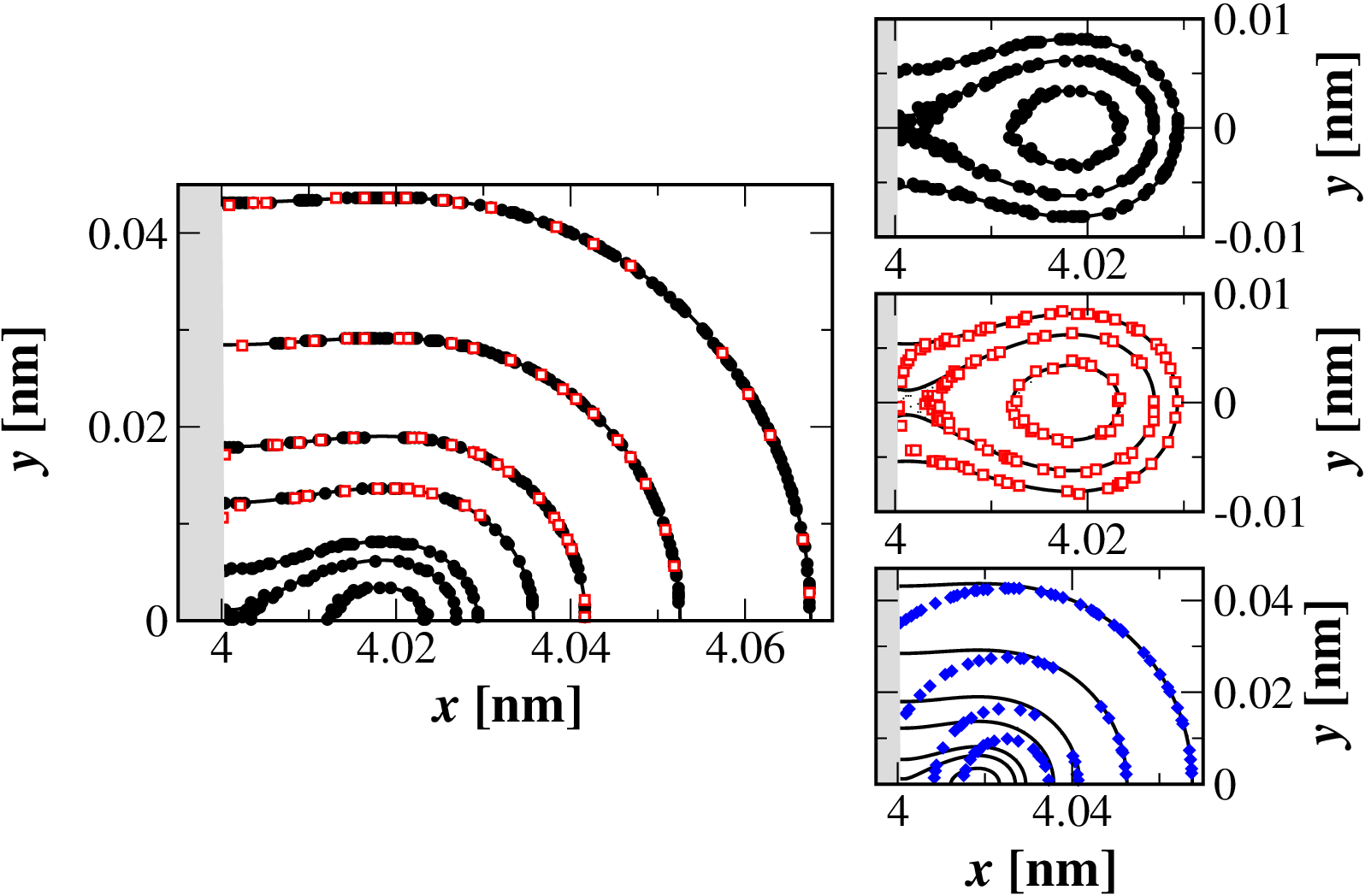}
\caption{\label{bild:diffusionAroundSphere} Contour plot of the
  transition probability of the relative vector of two proteins. Its
  evolution corresponds to a point particle with
  $D=D_1+D_2=0.24$\,nm$^2$/ns. The relative coordinate is initially
  $\vektor{r}(0)=(4.025\,\mbox{nm},0\,\mbox{nm})$ and evolves for the
  time $t = 0.001$\,ns. The force (for the relative component, compare
  relation (\ref{eq:relativeMotion})) is obtained from
  (\ref{eq:colloidPotential}) with $K=4\,\kbT$\,nm, $\kappa =
  0.02$\,nm$^{-1}$, $\sigma = 3.6$\,nm, $A = 3\,\kbT$ and $\delta =
  0.4$\,nm (compare e.\,g. Ref. \onlinecite{Pellicane2004}) making up
  an excluded hard-core region of radius $R_\indexHC=4$\,nm.  The
  numerical results are obtained from one integration step
  (i.e.~$\dx t = 0.001$\,ns) for which the encounter frequency is
  approximately 12.7\%. Exact curves obtained by solving the
  Fokker-Planck equation are displayed by solid lines. Left panel:
  (black) filled circles for our new algorithm, open (red) squares
  for event-driven scheme, for symmetry reasons only positive $y$ are
  shown. For the inner contour levels results from our method are
  shown separately in the upper right panel, those from
  the event-driven are shown in the middle panel on the right.
  The lower right panel exhibits the same data for the
  rejection scheme (blue filled diamonds). The grey shading represents
  the region which is inaccessible to the relative coordinate.}
\end{center}
\end{figure}

For two proteins that experience only their mutual interactions, the
motion of the common center-of-friction completely decouples from
their relative motion. Let us first look a this relative motion, which
then corresponds to a driven diffusive motion of a point-particle
around a sphere.  The transition probability for the relative
coordinate in the vicinity of the excluded hard-core region is shown
in Fig.  \ref{bild:diffusionAroundSphere}.  The time step is chosen to
be $\dx t = 0.001$\,ns so that $\mathcal{O}\left(\sqrt{\langle
  \vektor{r}^\power\rangle-\langle \vektor{r}\rangle^2}\right)\sim
0.02\mbox{\,nm}$. Therefore, we expect that the curvature of the
excluded hard-core sphere does not lead to considerable systematic
deviations from the exact transition probability.
The numerically obtained transition probabilities for one integration
step are compared with the 'exact' transition
probability that has been calculated by solving the corresponding
Fokker-Planck equation. 

\subsection{Mean first passage time}

As an example of a derived quantity we look more closely at the mean
first passage time for two different driven diffusive processes where
hard-wall interactions play an important role (see also
Fig. \ref{bild:passageTime} for a first brief example).  Although
these cases capture typical situations one encounters in simulations
of soft matter systems, they are simple enough so that the mean first
passage time obtained from the numerical integration of the equations
of motion can be compared with an analytical solution. The latter is
obtained from the mean escape time from an interval $[a,b]$ with a
reflecting boundary at $a$ and an adsorbing boundary at $b$,
\begin{equation}
\label{eq:escapeTimeFormula}
\tau(q_0) = \frac{1}{D} \int\limits_{q_0}^b \dx q \exp(\beta \Phi(q)) \int\limits_a^q \dx q' \exp(-\beta \Phi(q'))
\end{equation}
for a particle at inverse temperature $\beta = 1/\kb T$ moving in a potential
$\Phi(q)$, and starting
at $q_0$. \cite{Gardiner1983} 

\begin{figure}[h!]
\begin{center}
\includegraphics[scale=0.3,angle=0]{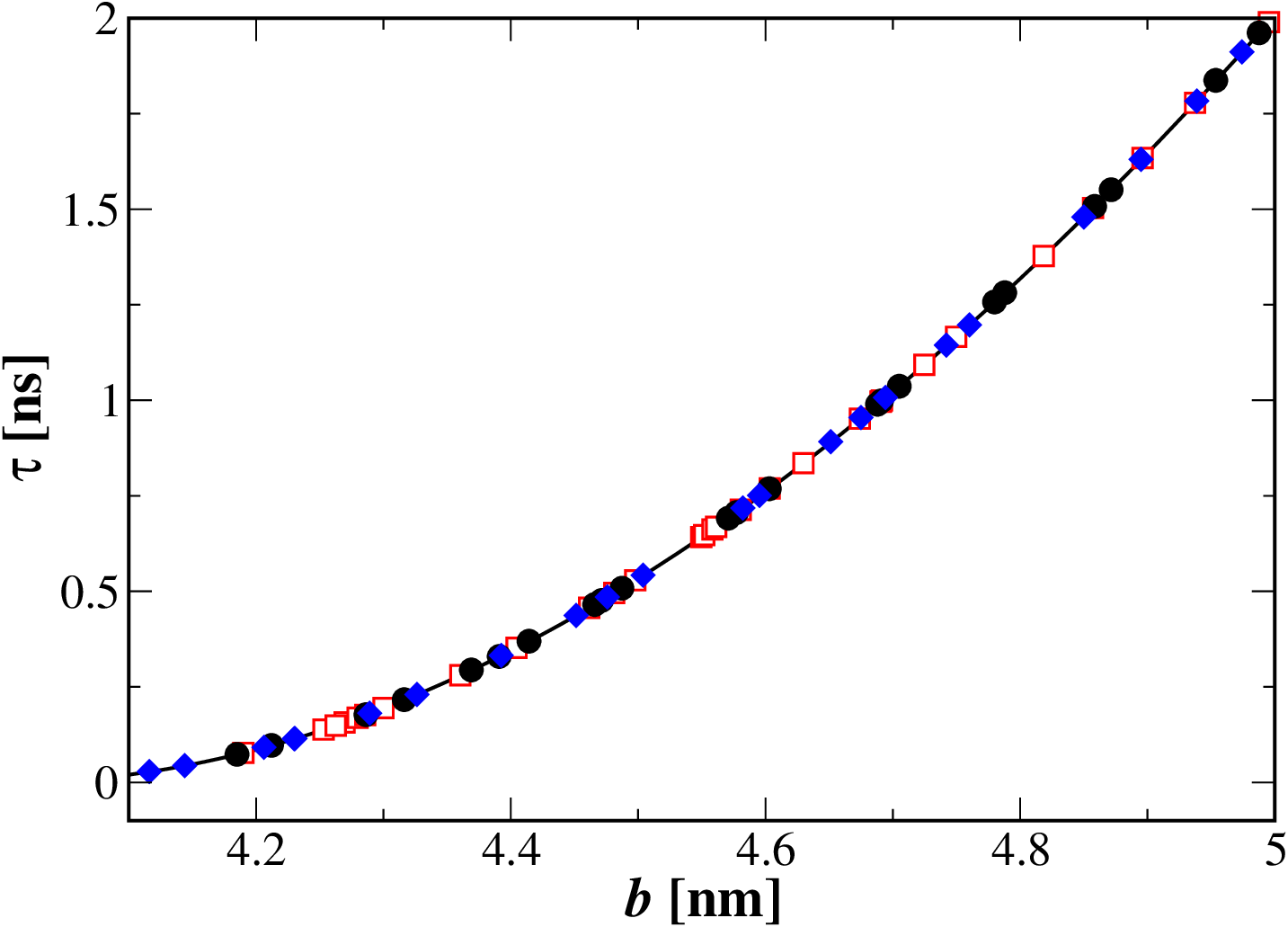}
\caption{\label{bild:escapeTime} Escape time $\tau$ from the potential
  (\ref{eq:colloidPotential}) as a function of the location $b$ of the
  (radial) threshold. The same potential parameters as in Fig.
  \ref{bild:diffusionAroundSphere} are used with $\vektor{r}(0) =
  (4.025\,\mbox{nm},0\,\mbox{nm})$ and $\dx t = 0.001$\,ns. With
  increasing $b$ the percentage of time steps that involve a 'collision'
  with the hard-core region drops from approximately
  13\% to 2\% for the shown data. The numerical data is shown by
  (black) circles for our proposed method, by (red) open squares for
  the event-driven scheme and by (blue) diamonds for the rejection
  scheme, the solid curve represents the exact escape time evaluated
  from (\ref{eq:escapeTimeFormula}) with $a=R_\indexHC = 4\,\mbox{nm}$, $q_0=4.025\,\mbox{nm}$
  and $\Phi(q) = \phi(q) - \kb T \ln r$
  (see (\ref{eq:colloidPotential}) and main text).
  The mean escape time was
  measured from the simulations as described in
  Refs. \onlinecite{Honerkamp1990,Mannella1999}.  
  }
\end{center}
\end{figure}

In the first situation we continue the example from the previous
section and consider the mean time the two proteins need to depart a
certain distance $b$. This is related to a thermally activated process
where the reaction coordinate is the separation $r =
|\vektor{r}|=|\vektor{r}_2 -\vektor{r}_1|$. Taking the interaction
potential (\ref{eq:colloidPotential}) the problem is radially
symmetric suggesting a transformation to polar coordinates. Due to the
isotropy of the fluctuating forces in the Langevin equation the
corresponding random forces in orthogonal polar coordinates are
uncorrelated. Therefore, the escape time is only determined by the
evolution equation for the separation $r$ which can be obtained from
(\ref{eq:relativeMotion}) by applying It\^o's formula
\cite{Gardiner1983}:
\begin{equation}
  \dx r = -\frac{1}{\eta_r} \frac{\dx}{\dx r} \Phi(r) \dx t + \sqrt{2D_r} \dx W_r
\end{equation}  
where $\eta_r = \kb T/D_r$ is the friction associated with the
relative coordinate ($D_r = D_1+D_2$) and $\dx W_r$ is an incremental
Wiener process. The effective potential is given by $\Phi(r) = \phi(r)
- \kb T \ln r$ in two space dimensions.  Figure \ref{bild:escapeTime}
depicts the exact values of the escape time as a function of the
threshold $b$ together with the numerically obtained values.  Before
the particle crosses the threshold at $b$ many integration steps have
to be carried out. In addition, a small time step $\dx t$ has to be
chosen to capture the variations in the force field so that only a
small fraction of encounter events will occur. It is thus expected
that the heuristic algorithms will produce good results. Figure
\ref{bild:escapeTime} indeed confirms this expectation. This example
points out that the rejection scheme may produce accurate results for
certain observables. However, this does not necessarily mean that the
rejection procedure is reliably applicable in general situations, as
already demonstrated with the examples above, and as will be
underlined by the following findings for the mean first passage time
of a particle close to a wall.

\begin{figure}[h!]
\begin{center}
\includegraphics[scale=0.3,angle=0]{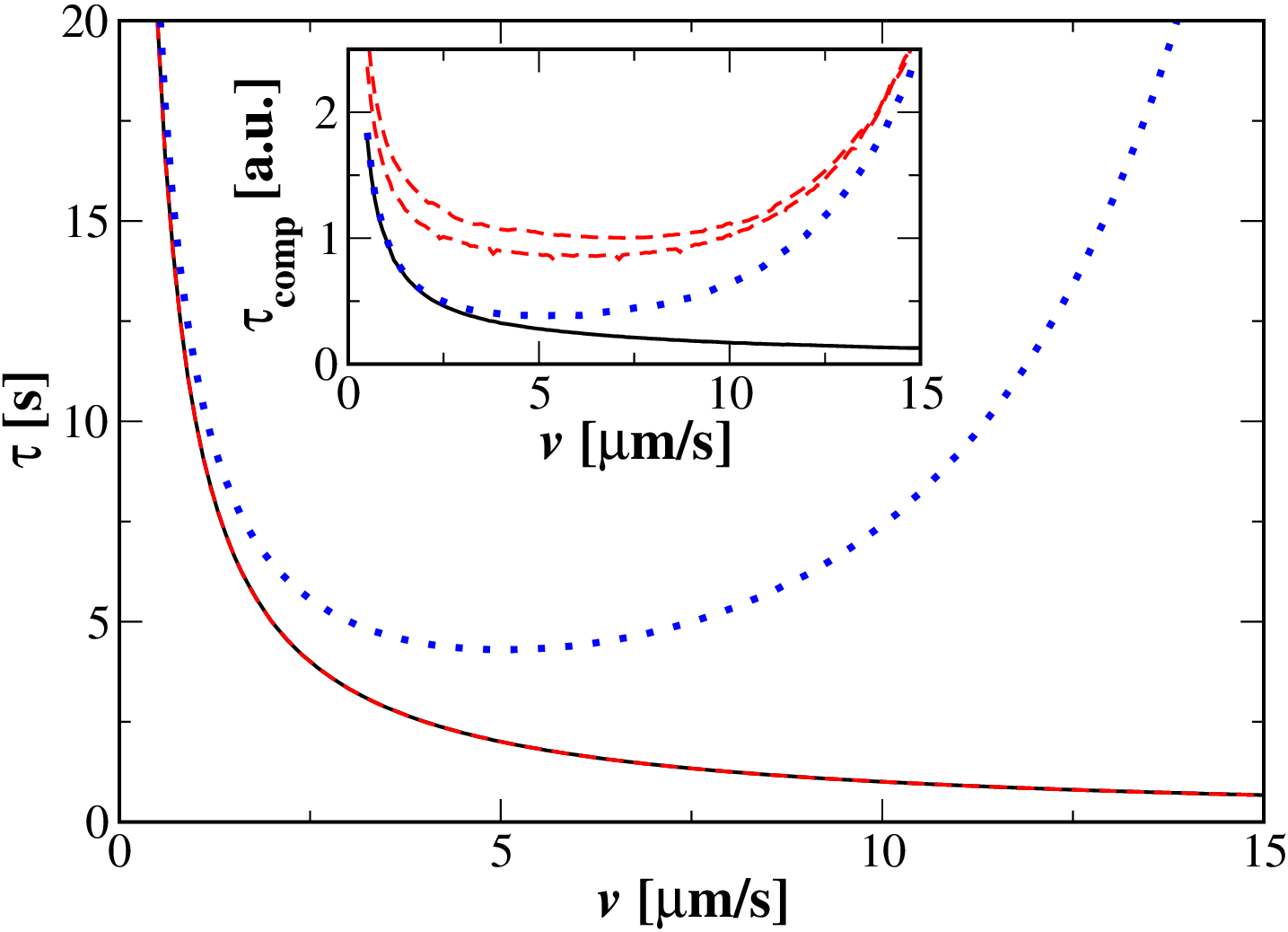}
\caption{\label{bild:escapeTimeWall} Mean first passage time of a
  particle (radius $1\,\mu\mbox{m}$) close to a wall (located at
  $x=0$) as a function of $v=f/\eta$. It moves from
  $\vektor{r}(0)=(0, 2\,\mu\mbox{m})$ to the line $y= b = 10\,\mu\mbox{m}$
  driven by the force $\vektor{F}=(-f,f)$ (see main text).  (For fixed
  $f$ the convergence as a function of $\dx t$ is similar to Fig.
  \ref{bild:passageTime}, not shown.)  The numerical data is shown by
  the (black) line for our method, which is indistinguishable from the
  exact passage time $\tau=b/v$ evaluated from
  (\ref{eq:escapeTimeFormula}) with $a=-\infty$, $b=10\,\mu\mbox{m}$,
  $q_0=0$ and $\Phi(q) = -fq = - \eta v q$.  The (red) dashed line
  represents the event-driven scheme and the (blue) dotted line the
  rejection scheme (time step $\dx t=0.01\,$s in all cases).  The mean
  escape time was measured from the simulations as described in
  Refs. \onlinecite{Honerkamp1990,Mannella1999}.  The inset shows the
  computation time consumed by the different algorithms.  (See Sec.
  \ref{sec:efficiency} on how the collision time is determined for the
  event-driven scheme. Upper dashed curve: stepwise propagation; lower
  dashed curve: nested intervals.)  For larger driving force, the
  computation times of the rejection and event-driven scheme is an
  order of magnitude longer as compared to our algorithm. For the
  shown data the 'collision' frequency increases from a few per cent
  to around 90\%.}
\end{center}
\end{figure}

As a second example we consider a (colloidal) particle which hits a
hard wall (for instance, a wall or obstacle in a microfluidic device
or an organelle in a cell etc.). The particle is driven by a
deterministic force with a non-vanishing component perpendicular to
the wall (for example due to an electric or magnetic field).  For
simplicity, we restrict ourselves again to two dimensions and assume
that the wall located at $x=0$ is infinitely extended along the $y$
direction such that it confines the particle to $x>0$ and $y \in
\mathbb{R}$. The particle is subject to the force
$\vektor{F}=(-f,f)$. In Fig. \ref{bild:escapeTimeWall} we show the
mean time a particle of radius $1\,\mu\mbox{m}$ starting at
$\vektor{r}(0)=(0, 2\,\mu\mbox{m})$ needs to 'slide along the wall' by
a distance of 10 particle radii until it crosses the line at
$b=y=10\,\mu\mbox{m}$. The results obtained for the different
integration schemes are compared to the exact solution calculated from
(\ref{eq:escapeTimeFormula}) with $\Phi(q) = -fq$ to be  $\tau=b\eta/f$.
While our new integration scheme yields the exact passage times, the
rejection scheme dramatically fails for too large $f$ as it
artificially slows down the particle motion due to prevailing rejection
steps until it practically stops completely. This goes in hand with a
considerable increased computing time, see inset of Fig.
\ref{bild:escapeTimeWall}. The event-driven scheme produces very good
results.  However, it requires (exponentially) increasing computation
time in the case of high collision frequency and strong forces into
the wall as many realizations of the random force for the propagation
step away from the wall will produce invalid end positions.

\subsection{Computational efficiency}
\label{sec:efficiency}

 To analyze the computational efficiency of our algorithm in
comparison to the rejection and event driven scheme in some more
detail and to get an impression of its performance in realistic
systems, we come back to the above example of the two interacting proteins.
As in realistic applications, we simulate the evolution of the
individual coordinates of the two interacting particles, i.e.\ we do
not decouple the center-of-friction motion. The event-driven scheme
requires the determination of the collision time of the two spheres
once an overlap has been detected. This collision time can be
determined analytically for this special case by solving a quartic
equation for non-vanishing forces (for zero force it reduces to a
quadratic equation). Having the possible generalization of an
arbitrarily shaped rigid particle in mind, however, we calculated the
collision time by an iterative scheme which was used by Tao et
al.\cite{Tao2006} They determined the collision time by subdividing
the integration step, which led to a particle overlap, into 100
sub-steps to successively propagate the particles along the ``Euler
path'' until the collision is detected. Alternatively , the collision
time can be determined by using nested intervals. Our method also
requires the (numerical) evaluation of a root when determining the
modified normal component of the relative motion (see Appendix \ref{sec:generateRandomQ}).
However, the motion perpendicular to the normal component and, in
particular, the evolution of the center coordinate need not be altered
and are indeed kept unchanged by the algorithm, so that only a
one-dimensional motion is affected. We therefore expect our algorithm
to be faster than the event-driven scheme where all components are
simultaneously affected when accounting for hard bodies. For the
typical parameters of the exemplary protein system from above, see for
example Fig. \ref{bild:diffusionAroundSphere}, we indeed found that
the required computing time of our method is smaller than the time for
the event-driven scheme. Not surprisingly, the discrepancy increases
for increasing encounter frequencies $w$. For the data presented in
this work we typically observe a difference of 10 \% to 30\% in
computational speed. The rejection scheme, on the other hand, is
typically 20\% faster at identical time steps.  However, to diminish
the systematic errors due to encounter events with hard walls the time
step has to be reduced accordingly by at least one magnitude, thus
requiring considerably increased computation time.

\begin{figure}[h!]
\begin{center}
\includegraphics[scale=0.325,angle=0]{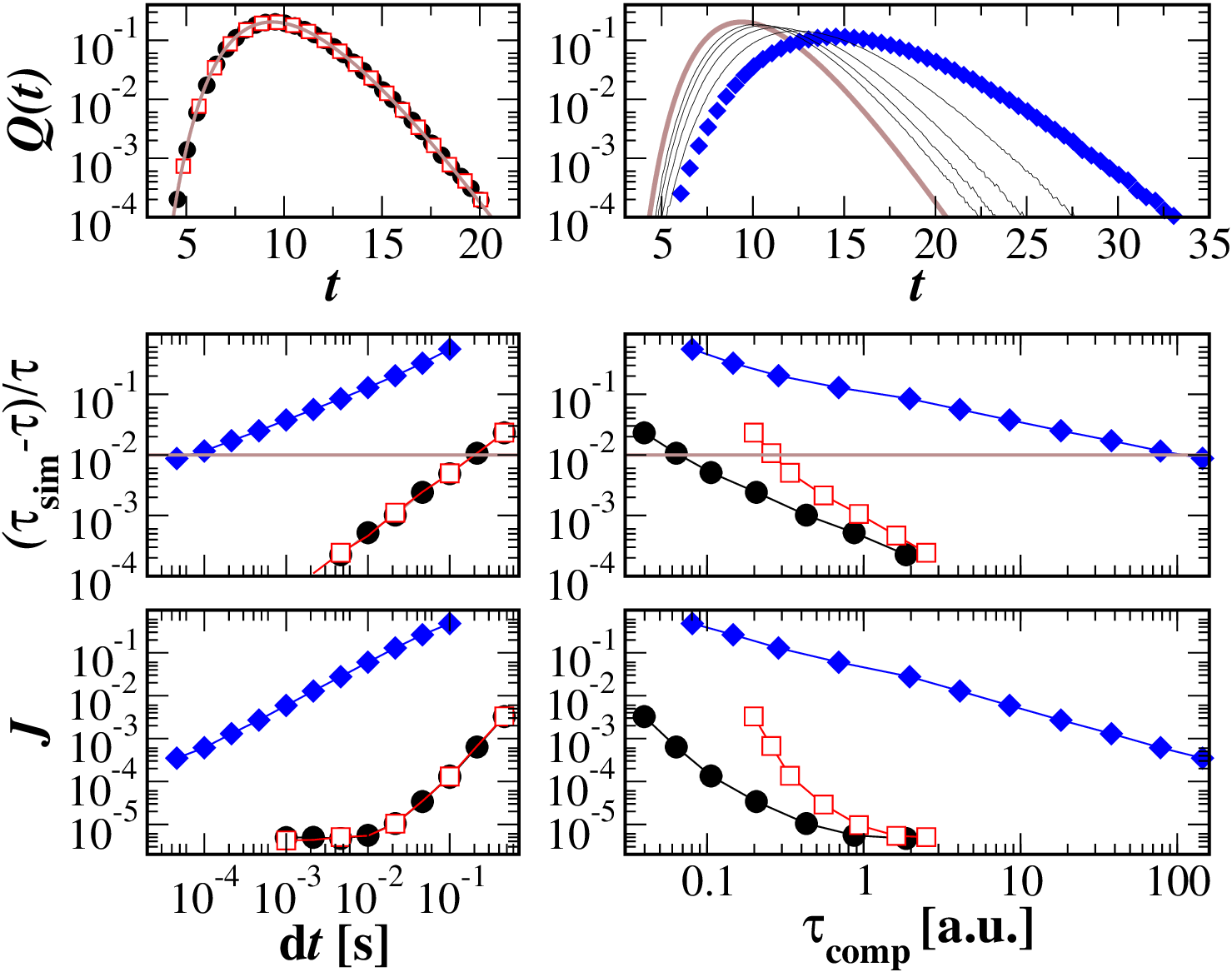}
\caption{\label{bild:jensenShannonAtWall} First passage time
  distribution for the system described in
  Fig. \ref{bild:escapeTimeWall} with $v=1\,\mu\mbox{m/s}$. Upper
  left panel: logarithm of the simulated distribution of our method
  (black circles) and of the event driven scheme (open red squares)
  together with the exact curve (solid brown line) for $\dx t = 0.1$
  s; upper right panel: same comparison for the rejection scheme
  (blue diamonds) and in addition curves for smaller $\dx t$ (thin
  black lines, $\dx t = 10^{-4/3}$, $10^{-5/3}$, $10^{-2}$,
  $10^{-7/3}$ s).  The middle panels show the relative deviation of
  the simulated mean first passage time $\tau_\indexSim$ from the
  exact value $\tau = 10$ s, the lower panels show the Jensen-Shannon
  divergence (\ref{eq:jensenShannon}). Both quantities are shown in double-logarithmic depiction
  as a function of the time step (left hand side) and the computation
  time (right hand side, the number of realizations is fixed to
  $10^7$, $\dx t$ is varied, nested intervals are used for the
  event-driven method). Note the different order of magnitude for the
  different methods (our method: black circles; event driven: open red
  squares; rejection: blue diamonds). The horizontal brown line
  indicates a prescribed accuracy of $(\tau_\indexSim-\tau)/\tau = 1$
  \%. For this setting the event driven-scheme exhibits a rather large
  encounter frequency of about 50 \% with the wall and needs by a
  factor three more computation time than our methods.  For $\dx
  t=0.01$ s (compare Fig. \ref{bild:escapeTimeWall}) an accuracy of
  $(\tau_\indexSim-\tau)/\tau \approx 0.5$ \% is achieved, the
  event-driven scheme then produces an encounter frequency of 11 \%
  and needs roughly 55\% more computation time. In contrast, the
  rejection scheme shows a very slow 'convergence'.  }
\end{center}
\end{figure}

For a more quantitative analysis we investigate the necessary
computational efforts to achieve a prescribed accuracy. To this end we
again use the passage time problem of a colloid gliding along a wall
of the previous subsection as for this example analytical results can
be obtained. For the system described in
Fig. \ref{bild:escapeTimeWall} (starting point $y_0 = 0$) the first
passage time distribution $Q(t)$ can be calculated\cite{Karatzas1988}
to be
\begin{equation}
\label{eq:fisrtPassageTimeDist}
  Q(t) = \frac{|b|}{\sqrt{4\pi Dt^3}}\exp\left( - \frac{(b-vt)^2}{4Dt}\right).
\end{equation} 
In addition to analyzing the relative deviation of the simulated mean
first passage time $\tau_\indexSim$ from the analytical result $\tau = b\eta/f$ we will
compare this exact distribution $Q$ with the simulated curve $Q_\indexSim$
by determining the Jensen-Shannon divergence\cite{Lin1991} as a
measure for the distance between $Q$ and $Q_\indexSim$. The
Jensen-Shannon divergence is defined by
\begin{equation}
\label{eq:jensenShannon}
  J = \frac{1}{2\ln 2}\int\limits_0^\infty \dx t \left( Q(t) \ln \frac{Q(t)}{M(t)}   + Q_\indexSim(t) \ln\frac{Q_\indexSim(t)}{M(t)}\right)
\end{equation} 
where $M = (Q + Q_\indexSim)/2$. $J$ can take on values between 0 and
1 with 0 indicating identical distributions.\cite{jensonShannonRemark}
The findings for the system of Fig. \ref{bild:escapeTimeWall} are
shown in Fig. \ref{bild:jensenShannonAtWall}. Surprisingly the
event-driven scheme produces results of the same accuracy as our new
methods for the same time step. Therefore, the computational
difference is solely determined by the effort for a single integration
step which amounts to typically 10-30 \% for moderate encounter
frequencies with the wall smaller than 10 \%. In accordance with our
general observation the data in Fig. \ref{bild:jensenShannonAtWall}
point out again that a rather small time step has to be chosen for the
rejection scheme to match the accuracy of the other methods.

\section{Conclusions and outlook}

The explicit form of the transition probability for a Brownian
particle in a constant force field on the one-dimensional half-line
shows that multiple encounters with hard bodies may lead to
contributions to that transition probability that cannot be generated
by a direct numerical integration of the Langevin equations. This results in
unsystematic and uncontrolled errors in heuristic
integration schemes, even for very small time steps.  For an algorithm
whose accuracy is to be systematically controlled by the size of the
discretization time step, the recourse to the transition probability
seems unavoidable.  

In this article we have developed a novel algorithm to account for
hard-body interactions in Brownian dynamics simulations with
controllable discretization error, and compared its performance with
heuristic schemes from the existing literature.  The central idea is
that, once a hard-body interaction has been detected during the
numerical integration of the Langevin equations of motion, the
component of the particle displacement(s) involved in the 'collision'
is generated directly from the solution for the transition probability
in the presence of a hard reflecting wall.  We demonstrated that the
aforementioned transition probability for the one-dimensional
half-line can be used to construct an approximate solution for general
many-dimensional physical systems with hard-body interactions.  The
algorithm decomposes the particle motion into parts that are
unaffected by the hard-body interaction and into an 'affected' part
along which the hard-wall 'collision' occurs.  The latter corresponds
to the relative motion of the collision partners and is reduced a
one-dimensional motion on a half-line with a hard, reflecting boundary
at the origin (the point of 'collision').  For this setting, the
general time-dependent transition probability is known analytically,
and can be used to generate relative displacements which are
consistent with the hard-body interactions and with the statistical
properties of the underlying stochastic evolution.

This principle can be extended to more general situations, for
example, where structured particles with a non-spherical shape are
present.  In general, the decomposition of the full particle
displacement(s) has to be adopted in such a way that the components
'unaffected' by the hard-body interaction and the 'affected' component
statistically decouple from each other (for the considered time step),
so that the hard-body 'collision' can be corrected without interfering
with the rest of the particle motions.  The 'unaffected' displacement
is thus retained (to save computation time) and together with the
corrected 'collision' part is used to reconstruct the overall
displacements of the involved particles. In this work we developed
this procedure for the case of interacting spherical particles, the
generalization to more complicatedly structured rigid particles will
be presented in a separate article. Furthermore, situations where
multiple particle overlaps occur, for example in dense systems, and
where secondary overlaps due to the applied modifications show up will
also be treated elsewhere.

\acknowledgements

Financial support of the SFB 613 and the SFB 625 is gratefully
acknowledged. We thank the experimental biophysics group of the
Bielefeld University for providing computational resources.

\appendix

\section{Generation of random numers distributed according to (\ref{eq:distributionOutwardWall})}
\label{sec:generateRandomQ}

One of the central building blocks of the described algorithm is the
generation of a random number $q$ that is distributed according to
(\ref{eq:distributionOutwardWall}). This can be achieved by using a
general transformation method (see
e.\,g. \onlinecite{Honerkamp1990} or \onlinecite{Press2007}). Consider a distribution function $p$
and define $F(q) = \int_0^q \dx q' p(q')$. If the number $x$ is
uniformly distributed on $[0,1]$ then $ q = F^{-1}(x)$ is distributed
according to the distribution $p$. Here $F^{-1}$ is the inverse of
$F$.

For the distribution function $p(q) = p_\indexHB(q,q_0,\dx t,v,
D_q)$ with $p_\indexHB$ given in (\ref{eq:distributionOutwardWall})
the corresponding integral is
\begin{eqnarray}
F(q) = \frac{\erfc\left( \frac{q_0 + v\dx t}{\sqrt{4D_q\dx t}}\right) -
  \exp\left(\frac{vq}{D_q}\right)\erfc\left(\frac{q+q_0+v\dx t}{\sqrt{4D_q\dx t}}\right)}
  {\erfc\left( \frac{q_0 + v \dx t}{\sqrt{4D_q\dx t}} \right)}.
\end{eqnarray} 
Due to the monotony of $F$ the equation $x=F(q)$ has precisely one
solution $q$ for given $x$. As to the best of our knowledge there is
no closed expression for the inverse of $F$, $q$ is obtained by
solving $x = F(q)$ numerically using, for example, Newton's methods,
bisection methods or combinations thereof. In this work Brent's scheme
\cite{Press2007} was used from the GNU scientific library \cite{GSL}.

\section{Harmonic oscillator on the half-line}
\label{sec:OU}

Let us consider a Brownian particle on a half-line in a harmonic
potential with spring constant $k$, i.\,e. the total potential is
given by
\begin{equation}
\label{eq:harmonicPotential}
\phi(q) =  \left\{ \begin{array}{ll}
         \frac{1}{2}k q^2, & \mbox{if $q \geq 0$}\\
        \infty, & \mbox{if $ q < 0$}\end{array} \right. 
\end{equation}
with $k>0$.  Introducing the characteristic time $\tau = \eta/k$ the
corresponding overdamped Langevin equation reads
\begin{equation}
\label{eq:harmonicLangevin}
\dx q = -\frac{q}{\tau}\dx t  + \sqrt{2D_q}\dx W
\end{equation}
with a reflective boundary at $q=0$. For the unrestricted harmonic
potential on the real axis the transition probability $p(q,t;q_0)$ of
the Ornstein-Uhlenbeck process (\ref{eq:harmonicLangevin}) is
analytically known\cite{Risken1984,vanKampen1987} for initial
condition $p(q,t=0;q_0) = \delta (q-q_0)$:
\begin{equation}
\label{eq:OUsolution}
p_\indexOU(q,t;q_0) = \frac{1}{\sqrt{2\pi\tau D_q (1-s^2)}} \exp\left( -\frac{(q-s q_0 )^2}{2\tau D_q (1-s^2) }  \right)
\end{equation}
where
\begin{equation}
\label{eq:sfu}
s= s(t) = \exp\left(-\frac{t}{\tau}\right) \, . 
\end{equation}
This solution can be used to construct the transition probability for the
harmonic oscillator (\ref{eq:harmonicPotential}) and reflecting
boundary condition on $\mathbb{R}_{>0}$ by using the image method:
\begin{equation}
\label{eq:OUwithWall}
p(q,t;q_0) = p_\indexOU(q,t;q_0) + p_\indexOU(-q,t;q_0) \, .
\end{equation}

\begin{figure}[h!]
\begin{center}
\includegraphics[scale=0.325,angle=0]{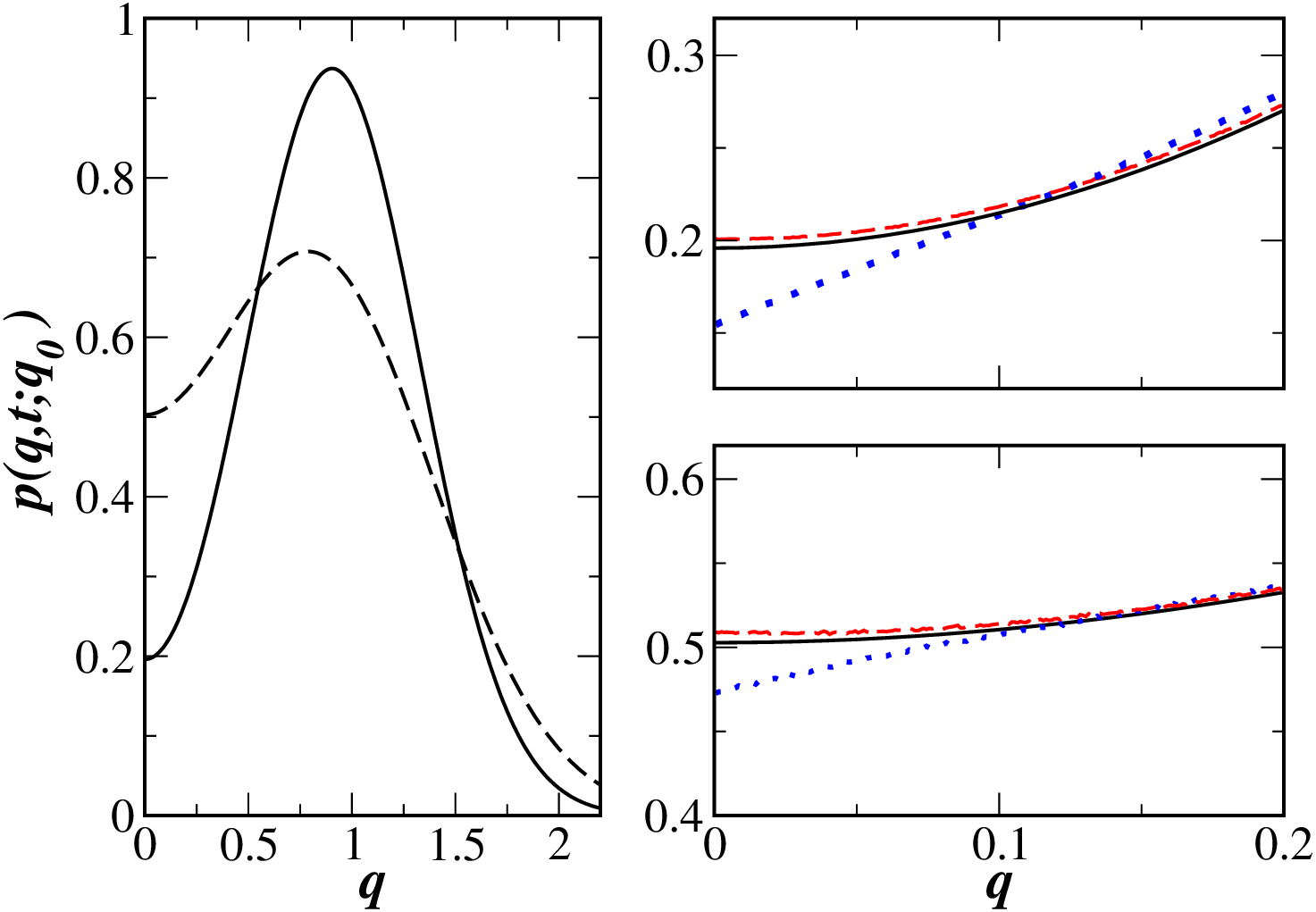}
\caption{\label{bild:onedimensionalOscillator} Transition probability
  of a particle on a half-line in the harmonic potential
  (\ref{eq:harmonicPotential}) for the parameters $\tau=1.0$, $D_q=1.0$
  and initial position $q_0=1.0$. In the left panel the analytical
  solution (compare (\ref{eq:OUwithWall}) and (\ref{eq:OUsolution}))
  is displayed for two times, namely $t=0.1$ (solid line) and $t=0.2$
  (dashed line). \cite{noteDimensionless} For the generation of the
  numerical data the rejection method and our scheme is used with a
  time step $\dx t = 0.01\tau = 0.01$. (The results from the
  event-driven algorithm are not displayed as they are indistinguishable
  from our method). The upper right panel shows a blow up for $t=0.1$,
  the lower one for $t=0.2$ (same magnification factor). For our new
  method (red dashed lines) there are still tiny deviations which are related to the
  finite discretization size
  of the time step. The rejection method (blue dotted lines) suffers from errors close
  to the wall due to the insufficient representation of hard-wall interactions.}
\end{center}
\end{figure}

Figure \ref{bild:onedimensionalOscillator} compares this exact
solution with the numerically obtained transition probability using
the algorithms discussed in the main text. To account for the
variations in the force field the integration time step is chosen to
be $\dx t = 0.01 \tau$. This time step is already small enough that
the solution from the event driven scheme is of similar numerical
accuracy as our new integration scheme.  For the rejection method,
however, there are still deviations close to the wall due to the
insufficient treatment of the reflecting boundary condition.

The transition probability of the unrestricted Ornstein-Uhlenbeck
process on $\mathbb{R}$ is a Gaussian. Therefore, an exact integration
algorithm for any $\dx t$ exists\cite{Gillespie1996}:
\begin{equation}
q(t+\dx t) = q(t) s(\dx t) + \sqrt{\tau D_q (1-[s(\dx t)]^2)} G
\label{eq:dqOU}
\end{equation}
with $G$ being a random Gaussian number of mean zero and variance one
and $s(\dx t)$ given in (\ref{eq:sfu}). Due to the symmetric structure
of the transition probability (\ref{eq:OUwithWall}) one can readily
construct an exact update scheme to generate valid positions $q^\ast$
for the potential (\ref{eq:harmonicPotential}) with a hard wall at the
center as well:
\begin{enumerate}
\item[1.]
Calculate a 'preliminary' position $q(t+\dx t)$ according to 
(\ref{eq:dqOU}) by a standard integration algorithm.
\item[2 a.]
If this proposed $q$ is in the physical domain, i.e. $q>0$, it is
accepted as the new position, $q^\ast=q$. 
\item[2 b.]
If this proposed position is unphysical, i.e. $q<0$, a new
physical position $q^*>0$ is obtained by simply reflecting $q$, i.e.\ $q^\ast = -q$.
\end{enumerate}

We finally remark that this procedure, the construction of valid end
positions by reflection, is possible because the exact solution can be
obtained by the image method. For $k\to 0$ the system reduces to free
diffusion on a half-line and the described method reduces to the one
discussed in Ref. \onlinecite{Scala2007} in the special case of one
space dimension.


\begin{thebibliography}{99}
\bibitem{fussnoteScala} {The scheme of Scala et al. \cite{Scala2007}
  can be developed systematically and envolves approximations which
  can be controlled within the numerics. However, this method works
  only for systems with vanishing (deterministic) forces.}
\bibitem{noteDimensionless} {As we do not have a particular system or physical context in mind, all parameters in Figs.
  \ref{bild:onedimensionalRejection}, \ref
      {bild:onedimensionalHeuristic}, \ref{bild:dispersionAtWall} and \ref{bild:dispersionAtWallDT} (as well as \ref{bild:onedimensionalOscillator})
      are given in terms of characteristic times, lengths, diffusion
      constants etc.}
\bibitem{noteMultipleCollisions}
Strictly speaking, we also have to assume that other hard boundaries or other particles,
which may be present in the system, are sufficiently far away
to not influence the considered displacement step (i.\,e. at least several standard deviations
of the typical displacement). Obviously, this cannot always be guaranteed, for example, at corners of
a structured boundary or for very dense suspensions.
An extension of our algorithm to include the case of 'multiple collisions'
is subject of current research.
\bibitem{jensonShannonRemark} { For the normal distribution
  $\mathcal{N}(\mu,\sigma^2)$ the two distributions $\mathcal{N}(0,1)$
  and $\mathcal{N}(0.1,1)$ have a Jenson-Shannon divergence 0.00125,
  the distributions $\mathcal{N}(0,1)$ and $\mathcal{N}(0.2,1)$ a
  divergence 0.005.  }
\end{thebibliography}
\end{document}